\documentclass{vgtc}                          % final (conference style)
\graphicspath{{figures/}{pictures/}{images/}{./}} % where to search for the images

\usepackage{times}                     % we use Times as the main font
\usepackage{tabu}                      % only used for the table example
\usepackage{booktabs}                  % only used for the table example
\usepackage{lipsum}                    % used to generate placeholder text
\usepackage{mwe}                       % used to generate placeholder figures

\usepackage{mathptmx}                  % use matching math font
\usepackage{amsmath}                   % for \text in math mode
\usepackage{amssymb}                   % for \mathbb
\usepackage{enumitem}                  % for compact itemize/enumerate

\usepackage[absolute,overlay]{textpos} % preprint
\usepackage[whole]{bxcjkjatype}
\onlineid{1549}

\vgtccategory{Research}

\vgtcinsertpkg

\preprinttext{To appear in an IEEE VGTC sponsored conference.}

\title{HaptoFlow: High-Fidelity Real-Time Vibrotactile Generation\\ via Flow Matching for Virtual Reality}

\author{Michikuni Eguchi\thanks{e-mail: eguchi@mvml.slis.tsukuba.ac.jp}\\ %
        \parbox{1.6in}{\scriptsize \centering University of Tsukuba \\ Metaverse Lab, Cluster, Inc.} %
\and Yuichi Hiroi\thanks{e-mail: y.hiroi@cluster.mu}\\ %
     \scriptsize Metaverse Lab, Cluster, Inc. %
\and Takefumi Hiraki\thanks{e-mail: hiraki@slis.tsukuba.ac.jp}\\ %
     \parbox{1.6in}{\scriptsize \centering University of Tsukuba \\ Metaverse Lab, Cluster, Inc.}}

\teaser{
  \centering
  \includegraphics[width=\linewidth]{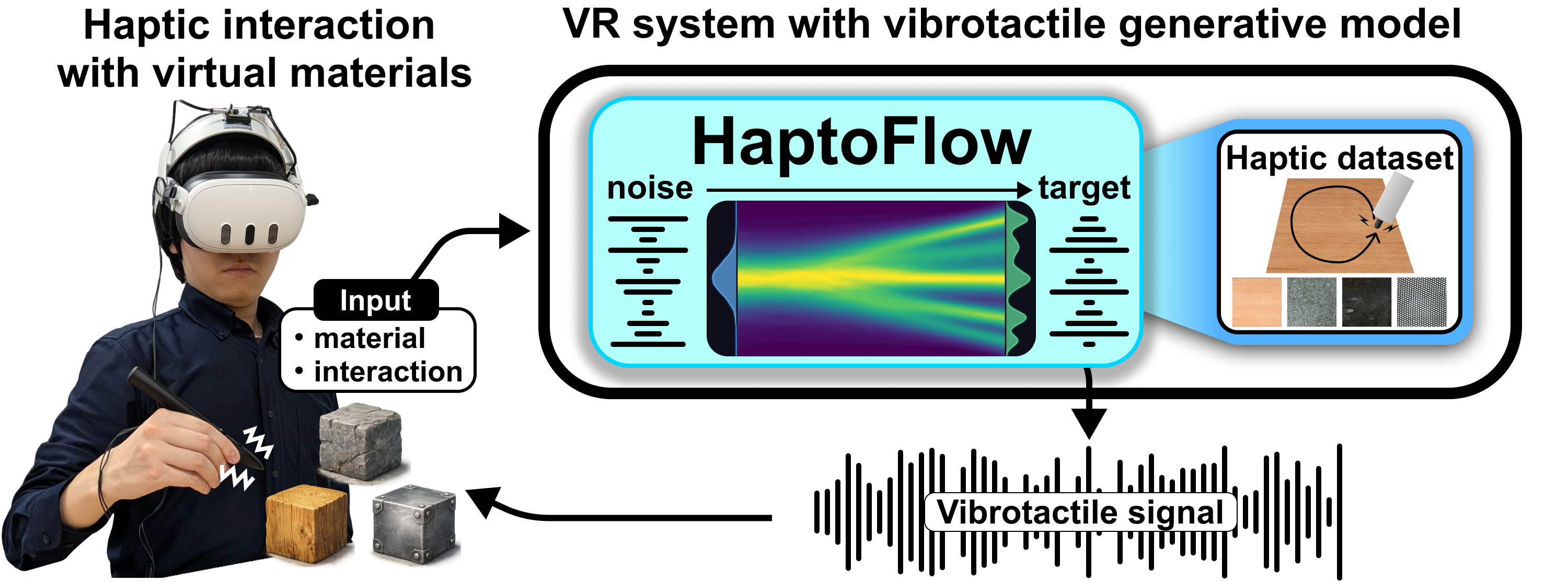}
  \caption{Overview of HaptoFlow and its integration into a VR system. HaptoFlow is a real-time vibrotactile generative model that produces realistic vibrotactile waveforms conditioned on material labels and interaction parameters (stroking velocity and applied force). Built on Flow Matching as its generative backbone, HaptoFlow outperforms existing models in both waveform reproduction accuracy and inference latency. This model enables scalable haptic design without manual authoring of individual waveforms.}
  \label{fig:teaser}
}

\abstract{
Haptic feedback is widely employed to enhance immersion in Virtual Reality (VR) environments. However, designing haptic stimuli that cover diverse interaction conditions remains a significant scalability challenge.
Data-driven haptic generation has emerged as a promising approach, yet existing models face an inherent trade-off between waveform expressiveness and inference responsiveness, which becomes increasingly critical as training data grow in scale and diversity.
To address this challenge, we propose HaptoFlow, a vibrotactile generative model based on Flow Matching, designed for interactive real-time haptic rendering in VR.
Flow Matching learns a continuous vector field that transforms a base distribution into the target data distribution, enabling efficient representation of complex haptic data distributions and thereby facilitating both high-quality generation and computational efficiency.
We train HaptoFlow conditioned on material labels and interaction parameters (stroking velocity and applied force), and integrate it into a VR system.
Technical evaluation demonstrates that HaptoFlow outperforms all baseline methods in both waveform reproduction accuracy and inference latency.
Furthermore, user studies confirm that the system latency falls well within the perceptual threshold of visual-haptic delay, and statistically significant improvements in perceived haptic quality are observed for a subset of materials.
These findings establish a practical foundation for scalable, data-driven haptic content creation in VR, and provide latency benchmarks that inform the design of future real-time haptic rendering systems.
Project page: \url{https://tamago117.github.io/HaptoFlow/}.
} % end of abstract


\AtBeginDocument{%
  \setlength\abovedisplayskip{5pt plus 1pt minus 1pt}%
  \setlength\belowdisplayskip{5pt plus 1pt minus 1pt}%
  \setlength\abovedisplayshortskip{2pt plus 1pt}%
  \setlength\belowdisplayshortskip{3pt plus 1pt minus 1pt}%
}

\begin{document}


\firstsection{Introduction}

\maketitle

% preprint 
\begin{textblock*}{170mm}(20mm,8mm)
\noindent\scriptsize © 2026 IEEE. Personal use of this material is permitted.
Permission from IEEE must be obtained for all other uses, in any current or
future media, including reprinting/republishing this material for advertising
or promotional purposes, creating new collective works, for resale or
redistribution to servers or lists, or reuse of any copyrighted component of
this work in other works.
\end{textblock*}

Enhancing immersion in Virtual Reality (VR) environments requires the complementary presentation of multiple sensory modalities tailored to the context of interaction.
Among these, haptic feedback plays an indispensable role in conveying the physical properties of objects, thereby heightening the sense of realism when users interact with virtual content~\cite{HapticVRReview2019}.
Vibrotactile feedback, in particular, has been widely adopted owing to its ease of implementation, with applications including the perception of surface roughness in VR environments~\cite{VibrotactileRoughness2024,VisuoTactileRoughness2025}.
To improve the quality of virtual experiences delivered through haptic stimulation, the design quality of haptic stimuli is therefore critical.
However, designing haptic stimuli often requires iterative refinement by domain experts~\cite{HapticExperienceDesign2017}, presenting a scalability challenge when deploying haptic feedback across diverse environments and interaction scenarios.
% JP: Virtual Reality（VR）環境における没入感の向上には，インタラクションの文脈に応じた複数の感覚モダリティの相補的な提示が必要である．
% JP: これらの中でも，触覚フィードバックは物体のテクスチャや硬さを伝える上で不可欠な役割を果たし，ユーザが仮想コンテンツとインタラクションする際のリアリティの感覚を高める~\cite{HapticVRReview2019}．
% JP: 特に振動触覚フィードバックは実装の容易さから広く採用されており，VR環境における表面粗さの知覚~\cite{VibrotactileRoughness2024,VisuoTactileRoughness2025}などへの応用がある．
% JP: 触覚刺激を通じて提供される仮想体験の品質を向上させるためには，触覚刺激のデザイン品質が極めて重要である．
% JP: しかしながら，触覚刺激のデザインには専門家による反復的な調整が必要となることが多く~\cite{HapticExperienceDesign2017}，多様な環境やインタラクションシナリオに触覚フィードバックを展開する際のスケーラビリティの課題を呈している．

Recent advances in generative AI have enabled the automation or assistance of design workflows across a wide range of domains, including image, audio, and text generation~\cite{stable-diffusion2022, le2023voicebox, yu2017seqgan}.
In the haptics domain, research on generating vibrotactile signals using data-driven models has been gaining momentum.
Models conditioned on designer-provided text or images have been proposed for vibrotactile signal generation~\cite{GANHaptics2021,HapticGen2025}.
Furthermore, methods that generate haptic signals in real time conditioned on continuously varying user interaction parameters have also been reported~\cite{transformer2022,DSTN2022,SPSI2024}.
The latter class of interaction-driven real-time generation is particularly promising for highly interactive VR environments.
However, the vibrotactile generative models proposed so far~\cite{transformer2022,DSTN2022,SPSI2024} are trained to directly regress a single waveform for a given set of conditioning inputs, which tends to average out the output and makes it difficult to faithfully reproduce the diverse behaviors of vibrotactile data that vary with material and interaction conditions.
Increasing model complexity to compensate for this limited expressiveness, in turn, raises inference latency, making the tradeoff between expressiveness and real-time responsiveness difficult to avoid.
Consequently, there remains a pressing need to develop haptic generative models that simultaneously achieve high expressiveness and low inference latency, enabling the practical deployment of generative haptic AI in diverse VR settings.
% JP: 近年の生成AIの進展により，テキストや画像生成~\cite{stable-diffusion2022, le2023voicebox, yu2017seqgan}をはじめとする幅広い分野でデザインワークフローの自動化や支援が可能になっている．
% JP: 触覚分野においても，データ駆動型モデルを用いた振動触覚信号の生成に関する研究が勢いを増している．
% JP: デザイナーが提供するテキストや画像を条件とした振動触覚信号生成モデルが提案されている~\cite{GANHaptics2021,HapticGen2025}．
% JP: さらに，連続的に変化するユーザインタラクション情報を条件としてリアルタイムに触覚信号を生成する手法も報告されている~\cite{transformer2022,DSTN2022,SPSI2024}．
% JP: 後者のインタラクション駆動型リアルタイム生成は，高度にインタラクティブなVR環境にとって特に有望である．
% JP: しかしながら，生成信号の表現力を向上させるためにモデルの複雑さを増すと推論レイテンシが上昇する傾向があり，表現力と応答性のトレードオフを回避することは困難である．
% JP: したがって，高い表現力と低い推論レイテンシを同時に達成し，多様なVR環境における生成触覚AIの実用的展開を可能にする触覚生成モデルの開発が依然として強く求められている．

To address this challenge, we propose HaptoFlow, a vibrotactile generative model based on Flow Matching~\cite{FlowMatching2023}, a generative modeling approach that learns a vector field governing the continuous transformation from a base distribution to the target data distribution.
Unlike conventional models that estimate a single deterministic output, generative modeling learns the conditional data distribution, enabling the generation of multiple plausible outputs even for data exhibiting diverse patterns or multimodal behavior that deterministic models tend to average out.
Flow Matching, in particular, represents complex data distributions as smooth continuous transformations, allowing the model to efficiently capture distributional structure and achieve both high generation quality and computational efficiency.
These properties have driven its adoption across a broad range of domains.
In particular, the ability to generate high-quality samples in as few as one or two ODE-solver steps~\cite{RectifiedFlow2023} makes Flow Matching well suited to real-time haptic rendering, where inference must complete within a few milliseconds.
% JP: 特に，わずか1〜2ステップのODEソルバで高品質なサンプルを生成できる能力~\cite{RectifiedFlow2023}は，推論を数ミリ秒以内に完了させる必要があるリアルタイム触覚レンダリングにおいてFlow Matchingを特に適したものとしている．
We build a Flow Matching-based vibrotactile generative model conditioned on material labels and interaction parameters (stroke velocity and applied force), and train it on real-world vibrotactile recordings.
We further integrate the proposed model into a VR pipeline combining a head-mounted display (HMD) and a stylus pen, constructing an end-to-end system capable of interactive real-time vibrotactile feedback, to demonstrate its practical applicability (\cref{fig:teaser}).
% JP: この課題に取り組むため，我々はFlow Matching~\cite{FlowMatching2023}に基づく振動触覚生成モデルHaptoFlowを提案する．Flow Matchingは基底分布から目標データ分布への連続的な変換を支配するベクトル場を学習する生成モデリング手法である．
% JP: 単一の決定論的出力を推定する従来モデルとは異なり，生成モデリングは条件付きデータ分布を学習するため，決定論的モデルが平均化してしまうような多様なパターンやマルチモーダルな挙動を示すデータに対しても，複数の妥当な出力を生成することが可能である．
% JP: 特にFlow Matchingは，複雑なデータ分布を滑らかな連続変換として表現するため，分布構造を効率的に捉えることができ，高い生成品質と計算効率の両立を実現する．
% JP: これらの特性により，Flow Matchingは幅広い分野で採用されている．
% JP: 我々は，素材ラベルとインタラクションパラメータ（ストローク速度と印加力）を条件とするFlow Matchingベースの触覚生成モデルを構築し，実世界の振動触覚録音データを用いて学習した．
% JP: さらに，提案モデルをヘッドマウントディスプレイ（HMD）とスタイラスペンを組み合わせたVRパイプラインに統合し，インタラクティブなリアルタイム振動触覚フィードバックが可能なエンドツーエンドシステムを構築した（\cref{fig:teaser}）．

We conducted three experiments to evaluate the proposed system:
(i) a technical evaluation comparing the proposed model against existing baselines in waveform reproduction accuracy and inference latency,
(ii) a user study investigating the perceptual threshold for visual-haptic delay when using a stylus pen in VR, and
(iii) a user study assessing the perceptual similarity between haptic stimuli generated by the proposed and baseline methods and those produced by real physical objects.
% JP: 提案システムを評価するために3つの実験を実施した：(i) 波形再現精度と推論レイテンシにおける既存ベースラインとの技術的比較，(ii) VRでスタイラスペン使用時のvisual-haptic遅延知覚閾値の調査，(iii) 提案手法・ベースライン手法の生成触覚刺激と実物体との知覚的類似性の評価．

In summary, the contributions of this paper are:
\begin{itemize}[topsep=2pt, itemsep=2pt, parsep=0pt]
  \item A Flow Matching-based vibrotactile generative model that achieves state-of-the-art waveform reproduction accuracy and inference latency compared to existing methods.
  \item A comprehensive evaluation of the proposed model, comprising technical benchmarks (waveform reproduction accuracy and inference latency) and user studies in an interactive VR system, assessing visual-haptics delay tolerance for a VR HMD with a stylus pen and the perceptual similarity of generated stimuli to real physical objects.
\end{itemize}
% JP: 本論文の貢献を以下にまとめる：
% JP: \begin{itemize}
% JP: 既存手法と比較して最先端の波形再現精度と推論レイテンシを達成する，Flow Matchingベースの振動触覚生成モデル．
% JP: 提案モデルの包括的な評価．技術的ベンチマーク（波形再現精度と推論レイテンシ）と，インタラクティブVRシステム上でのユーザスタディからなり，VR HMDとスタイラスペンにおけるvisual-haptics遅延許容量および生成刺激と実物体との知覚的類似性を評価する．
% JP: \end{itemize}

% Fig: アーキテクチャの図
\begin{figure*}[t]
    \centering
    \includegraphics[width=1\linewidth]{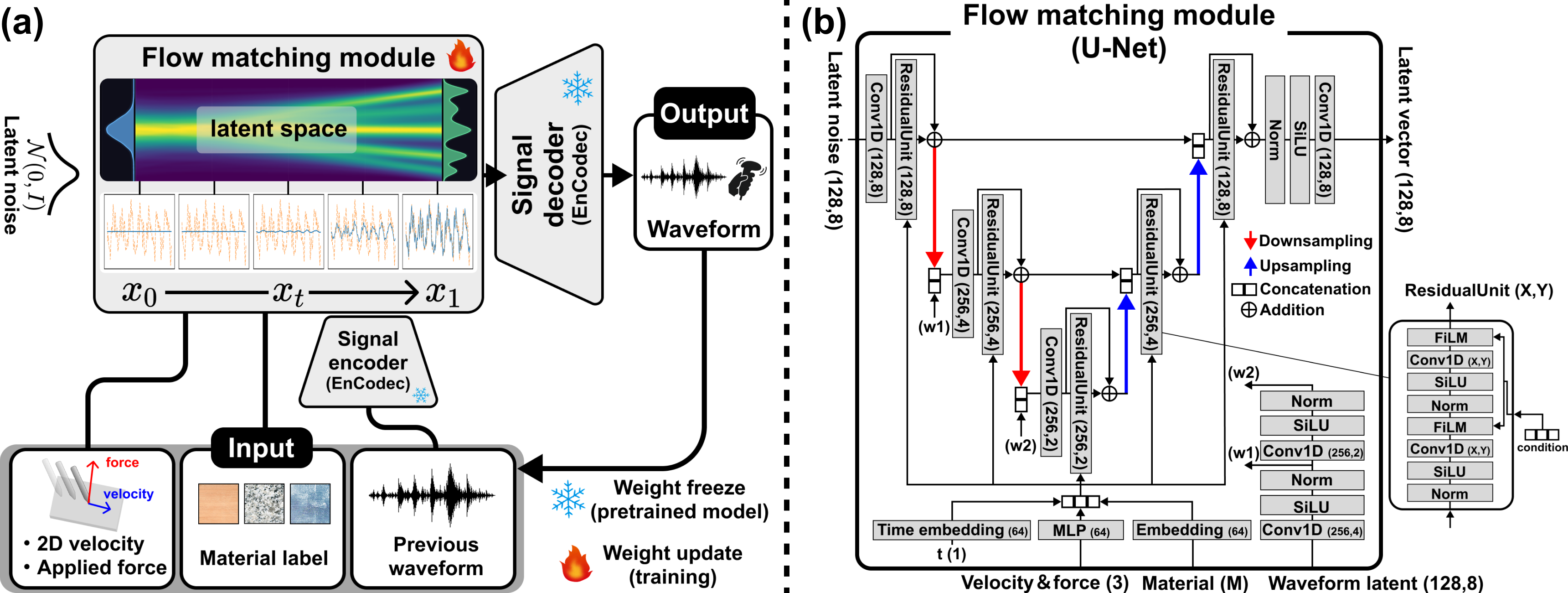}
    \caption{Architecture of the proposed model. (a) The model comprises a Signal Encoder-Decoder (EnCodec) and a Flow Matching module with a U-Net backbone. (b) Material labels and interaction parameters are injected via FiLM conditioning at every level of the U-Net.}
    \label{fig:architecture}
\end{figure*}

\section{Related Work}
% JP: （ロードマップ段落削除：サブセクション見出しで構造が自明なため）

\subsection{Haptic Feedback in VR}
\label{sec:rw-haptic-vr}
% VR環境における没入感の向上を目的として，力覚提示，温度提示，圧力提示など様々な触覚モダリティを活用した提示手法が研究されてきた．
% 力覚提示では，外骨格型やグローブ型デバイスを用いた仮想物体との接触力のフィードバックが実現されており~\cite{Dexmo2016,FluidReality2023,Wolverine2016}，温度提示ではペルチェ素子などを利用した仮想環境中の熱的特性の伝達~\cite{ThermoGrasp2024,ThermalDisplayGlove2020}など，多様なアプローチが探求されている．

% 振動触覚もそのような触覚モダリティの一つであり，既存のVRデバイスへの組み込みが容易であることから特に広く採用されている．
% 物体表面の粗さ知覚~\cite{VibrotactileRoughness2024,VisuoTactileRoughness2025}，物体操作時のフィードバック~\cite{VRTangible2024}，空間知覚の強化~\cite{SpatializedVibrotactile2022}，注意喚起~\cite{InvisibleBoundaries2020}といった多様な場面での活用が報告されており，盛んに研究が進められている。
% 一方で，多様な素材や環境に対応した触覚波形の設計には専門的な調整が必要であり~\cite{HapticExperienceDesign2017}，スケーラビリティが課題として残されている．
% この課題に対するアプローチとして，データ駆動型の触覚生成モデルの研究が近年進展しており，次節で詳しく述べる．

Among the haptic modalities explored to enhance VR immersion, including force~\cite{Dexmo2016, FluidReality2023, Wolverine2016} and thermal~\cite{ThermoGrasp2024, ThermalDisplayGlove2020} feedback, vibrotactile feedback has been particularly widely adopted owing to its ease of integration into existing VR devices.
% JP: VR没入感向上のために探求されてきた触覚モダリティ（力覚~\cite{Dexmo2016, FluidReality2023, Wolverine2016}や温覚~\cite{ThermoGrasp2024, ThermalDisplayGlove2020}フィードバックを含む）の中で，振動触覚フィードバックは既存VRデバイスへの統合の容易さから特に広く採用されている．
Its utility has been demonstrated across diverse scenarios, including the perception of surface roughness~\cite{VibrotactileRoughness2024, VisuoTactileRoughness2025}, object manipulation feedback~\cite{VRTangible2024}, spatial awareness enhancement~\cite{SpatializedVibrotactile2022}, and attention guidance~\cite{InvisibleBoundaries2020}.
Despite this breadth of application, designing haptic waveforms that faithfully represent diverse materials and contact conditions requires expert knowledge and laborious manual tuning~\cite{HapticExperienceDesign2017}, posing a scalability challenge.
Data-driven haptic generative models have emerged as a promising approach to address this challenge.
% JP: これらのモダリティの中で，振動触覚フィードバックは既存のVRデバイスへの統合が容易であることから，特に広く採用されている．
% JP: その有用性は，表面粗さの知覚~\cite{VibrotactileRoughness2024, VisuoTactileRoughness2025}，物体操作フィードバック~\cite{VRTangible2024}，空間認識の強化~\cite{SpatializedVibrotactile2022}，注意誘導~\cite{InvisibleBoundaries2020}など多様な場面で実証されている．
% JP: このような幅広い応用にもかかわらず，多様な素材や接触条件を忠実に表現する触覚波形の設計には専門知識と多大な手動調整が必要であり~\cite{HapticExperienceDesign2017}，スケーラビリティの課題を提起している．
% JP: データ駆動型の触覚生成モデルは，この課題に対処する有望なアプローチとして登場しており，次節で詳しくレビューする．

\subsection{Vibrotactile Rendering}
\label{sec:rw-haptic-model}

Vibrotactile rendering of material surfaces involves strong nonlinearity and highly diverse patterns depending on interaction conditions, motivating data-driven approaches from early work: Okamura et al.~\cite{RealityBasedModels2001} fitted recorded surface vibrations to exponentially decaying sinusoids, while Culbertson et al.~\cite{AutoregressiveHapticTextureModel2014} employed autoregressive models conditioned on velocity and applied force for real-time texture synthesis.
The systematic collection and release of haptic texture databases~\cite{LMTDatabase2014, LMTDatabase2017, AutoregressiveHapticTextureModel2014, RobustSurfaceRecognition2024, ClusterHapticTextureDataset2025} has further provided a shared foundation for training and evaluating such models.
% JP: 素材表面の振動触覚レンダリングは強い非線形性と多様なパターンを含み，初期の研究からデータ駆動型アプローチが動機づけられてきた：Okamuraら~\cite{RealityBasedModels2001}は記録された表面振動を指数減衰正弦波にフィッティングし，Culbertsonら~\cite{AutoregressiveHapticTextureModel2014}は速度と印加力を条件とする自己回帰モデルをリアルタイムテクスチャ合成に用いた．
% JP: 触覚テクスチャデータベースの体系的な収集と公開~\cite{LMTDatabase2014, LMTDatabase2017, AutoregressiveHapticTextureModel2014, RobustSurfaceRecognition2024, ClusterHapticTextureDataset2025}が，こうしたモデルの訓練・評価のための共通基盤を提供してきた．

These early approaches, however, fit a separate model for each material and rely solely on low-level interaction parameters, so they scale poorly as material sets expand and cannot accommodate higher-level inputs such as images or language.
With the rapid advancement of deep learning, neural network (NN)-based haptic generative models have emerged as a powerful alternative, representing diverse materials within a single network and admitting these richer conditioning inputs~\cite{HapticGen2025, Sound2Hap2026, GANHaptics2021}.
These models can be broadly classified into pre-generative and real-time generative models.
% JP: ただし，これらの初期手法は素材ごとに個別のモデルをフィッティングし，低レベルのインタラクションパラメータのみに依存するため，素材数の増加に対してスケールしにくく，画像や言語といった高レベルの入力を扱えない．
% JP: 深層学習の急速な進展に伴い，NNに基づく触覚生成モデルが強力な代替手法として登場し，多様な素材を単一のネットワークで表現し，こうしたより豊かな条件付け入力を受け入れられる~\cite{HapticGen2025, Sound2Hap2026, GANHaptics2021}．
% JP: これらのモデルは事前生成モデルとリアルタイム生成モデルに大別できる．
Pre-generative models synthesize haptic waveforms offline from text, audio, or images~\cite{HapticGen2025, Sound2Hap2026, GANHaptics2021}, but produce static stimuli that cannot adapt to runtime user behavior.
% JP: 事前生成モデルはテキスト・音声・画像から触覚波形をオフラインで合成するが~\cite{HapticGen2025, Sound2Hap2026, GANHaptics2021}，実行時のユーザ行動に適応できない静的な刺激を生成する．
Real-time generative models, in contrast, synthesize haptic waveforms dynamically in response to user actions during a VR session.
By conditioning waveform generation on parameters such as the material label of the contacted object, the tool velocity, and the applied force, these models can continuously adapt the haptic output to the evolving contact state~\cite{transformer2022, DSTN2022, SPSI2024}.
% JP: 一方，リアルタイム生成モデルは，VRセッション中のユーザの動作に応じて触覚波形を動的に合成する．
% JP: 接触した物体の素材ラベル，ツールの速度，印加力などのパラメータを条件として波形生成を行うことで，これらのモデルは変化する接触状態に触覚出力を継続的に適応させることができる~\cite{transformer2022, DSTN2022, SPSI2024}．

For interactive VR applications, real-time generative models are particularly promising; however, the generated waveforms must achieve high perceptual fidelity while keeping inference latency sufficiently low.
These requirements are in tension with data-driven scaling, since larger and more diverse datasets demand greater model capacity and thus longer inference time.
Furthermore, to the best of our knowledge, no prior work has integrated a data-driven haptic generative model into an interactive VR environment.
The present work therefore proposes a Flow Matching-based haptic generative model and demonstrates its integration into a VR system as an end-to-end interactive haptic rendering pipeline.
By advancing waveform fidelity and inference speed together, it also reinforces the core generation capability on which the input-flexible models above rely.
% JP: リアルタイム触覚レンダリングは高い波形忠実度と低い推論レイテンシの両立を求めるが，データセットの拡大とモデル容量の増大に伴い推論時間が増大する傾向があり，これらの要件は相反する．
% JP: 我々の知る限り，データ駆動型振動触覚生成モデルをインタラクティブなVR環境に統合した先行研究は存在しない．
% JP: 本研究では，Flow Matchingに基づく振動触覚生成モデルを提案し，エンドツーエンドのインタラクティブ触覚レンダリングパイプラインとしてVRシステムへの統合を実証する．
% JP: 波形忠実度と推論速度を同時に高めることで，本研究は上述の入力フレキシブルなモデルが依拠する中核的な生成能力を強化する．

\subsection{Generative Modeling}
\label{sec:rw-generative}

Generative models learn the underlying probability distribution of training data directly, enabling the synthesis of diverse and stochastic samples that are difficult to produce with conventional deterministic approaches.
Representative families include Conditional Variational Autoencoders (CVAEs)~\cite{CVAE2015}, Generative Adversarial Networks (GANs)~\cite{GAN2014}, Diffusion Models~\cite{DDPM2020}, and Flow Matching~\cite{FlowMatching2023}, all of which have been applied across image, audio, and text domains~\cite{stable-diffusion2022, le2023voicebox, yu2017seqgan}.
In the haptic domain, however, adoption of generative modeling has so far been limited, with GAN-based haptic generation~\cite{GANHaptics2021} and, more recently, diffusion-based audio-haptic generation~\cite{kotani2026audiohapticdiffusion} among the few reported examples.
% JP: 生成モデルは訓練データの背後にある確率分布を直接学習し，従来の決定論的アプローチでは生成が困難な多様かつ確率的なサンプルの合成を可能にする．
% JP: 代表的な手法群として，Conditional Variational Autoencoders (CVAEs)~\cite{CVAE2015}，Generative Adversarial Networks (GANs)~\cite{GAN2014}，Diffusion Models~\cite{DDPM2020}，およびFlow Matching~\cite{FlowMatching2023}があり，いずれも画像，音声，テキストの各領域に適用されている~\cite{stable-diffusion2022, le2023voicebox, yu2017seqgan}．
% JP: しかし，触覚領域における生成モデルの採用はこれまで限定的であり，我々の知る限り，GANに基づく触覚生成~\cite{GANHaptics2021}や，より最近ではdiffusionに基づく音響・触覚生成~\cite{kotani2026audiohapticdiffusion}が数少ない報告例である．

Among these, CVAEs offer stable conditional generation but tend to produce blurry outputs with limited high-frequency fidelity; GANs achieve sharper samples through adversarial training yet suffer from training instability and mode collapse; and Diffusion Models yield high-quality diverse samples but typically require 50--1000 denoising steps, incurring substantial inference cost.
% JP: これらのうち，CVAEsは安定した条件付き生成が可能だが高周波忠実度の低いぼやけた出力を生成する傾向があり，GANsは敵対的訓練により鮮明なサンプルを実現するが訓練不安定性やモード崩壊の問題があり，Diffusion Modelsは高品質で多様なサンプルを生成するが通常50〜1000ステップのノイズ除去を要し大きな推論コストを招く．

Flow Matching~\cite{FlowMatching2023} directly learns a vector field that defines a continuous transport from a base distribution to the target data distribution, and has attracted attention as an approach that overcomes the inference bottleneck of Diffusion Models.
By formulating training as regression of conditional vector fields, Flow Matching achieves stable learning while requiring only a small number of ODE-solver steps to generate high-quality samples at inference, and outperforms Diffusion Models in both training and inference efficiency~\cite{FlowMatching2023}.
% JP: Flow Matching~\cite{FlowMatching2023}は，基底分布から目標データ分布への連続的な輸送を定義するベクトル場を直接学習する手法であり，Diffusion Modelsの推論ボトルネックを克服するアプローチとして注目を集めている．
% JP: 訓練を条件付きベクトル場の回帰として定式化することで，Flow Matchingは安定した学習を実現しつつ推論時には少数のODEソルバステップのみで高品質なサンプルを生成し，訓練・推論の両効率でDiffusion Modelsを上回る~\cite{FlowMatching2023}．

Liu et al.~\cite{RectifiedFlow2023} formalized this efficiency by showing that straight-line ODE trajectories enable accurate generation in as few as one Euler step, a principle validated at scale in image synthesis~\cite{SD3_2024} and speech synthesis~\cite{MatchaTTS2024}.
Both domains share with haptic rendering the demand for high-fidelity continuous signals under tight inference budgets, yet Flow Matching has seen little use on vibrotactile signals.
To the best of our knowledge, the present work is the first to apply Flow Matching to temporal vibrotactile waveform generation conditioned on interaction parameters.
% JP: この効率性の理論的基盤はLiuら~\cite{RectifiedFlow2023}により形式化されており，ソース分布とターゲット分布を結ぶ直線的なODE軌道を学習することで，わずか1ステップのEuler法でも正確な生成が可能であることが示された．
% JP: この原理は画像合成において大規模に検証されており，Esserら~\cite{SD3_2024}はRectified FlowトランスフォーマーがDiffusion定式化を複数の指標で上回ることを実証した．
% JP: 画像以外でも，Flow Matchingは音声合成において最先端の結果を達成しており，Matcha-TTS~\cite{MatchaTTS2024}はわずか2〜4ステップのODEで高品質な出力を実現している．
% JP: これらの領域は，厳しい推論制約下で高忠実度の連続信号を生成するという要求を触覚レンダリングと共有しているが，Flow Matchingの振動触覚信号への適用はほとんど行われていない．
% JP: 我々の知る限り，本研究はFlow Matchingをインタラクションパラメータを条件とした時間的振動触覚波形生成に適用した最初の研究である．

These properties make Flow Matching well suited to real-time VR haptic rendering, where waveforms must be synthesized within a few milliseconds of user interaction.
We exploit this advantage and propose a Flow Matching-based vibrotactile generative model for interactive VR.
% JP: これらの特性により，Flow Matchingはユーザインタラクションから数ミリ秒以内に波形を合成する必要があるリアルタイムVR触覚レンダリングに適する．
% JP: 本研究では，この利点を活用し，インタラクティブVRのためのFlow Matchingベースの振動触覚生成モデルを提案する．

% visual-haptics delay
\subsection{Visual-Haptics Delay Tolerance of Human}
\label{sec:rw-delay}

Presenting visual and haptic stimuli in temporal synchrony is critical for VR user experience, and keeping this delay within a perceptually tolerable range is an important design constraint.
The tolerable delay varies depending on the haptic device and display system employed.
Miyatake et al.~\cite{HaptoMapping2023} reported detection thresholds of approximately 100~ms for a finger-mounted device, 160~ms for a stylus-type device, and 500~ms for an arm-mounted device in a tabletop projection display.
Similarly, Nagano et al.~\cite{HaptoFloater2024} reported a threshold of approximately 110~ms for a finger-mounted device in a mid-air image display.
Di Luca and Mahnan~\cite{PerceptualLimitsVR2019} reported a threshold of approximately 100~ms for a glove-type device in a VR environment.
% JP: 視覚刺激と触覚刺激を時間的に同期して提示することはVRユーザ体験にとって極めて重要であり，この遅延を知覚的に許容可能な範囲内に収めることは重要な設計制約である．
% JP: 許容遅延は使用する触覚デバイスやディスプレイシステムによって異なる．
% JP: Miyatakeら~\cite{HaptoMapping2023}は，卓上型プロジェクションディスプレイにおいて，指装着型デバイスで約100~ms，スタイラス型デバイスで約160~ms，腕装着型デバイスで約500~msの検出閾値を報告した．
% JP: 同様に，Naganoら~\cite{HaptoFloater2024}は空中像ディスプレイにおける指装着型デバイスで約110~msの閾値を報告した．
% JP: Lucaら~\cite{PerceptualLimitsVR2019}はVR環境におけるグローブ型デバイスで約100~msの閾値を報告した．

Although such thresholds have been characterized across various systems, to the best of our knowledge, no prior study has reported the tolerance for a VR HMD combined with a stylus pen.
To verify that the constructed VR system operates within the perceptually tolerable delay range, this paper also reports experimental results on visuo-haptic latency.
% JP: このような閾値は様々なシステムにおいて報告されているが，我々の知る限り，VR HMDとスタイラスペンの組み合わせに対する許容遅延を報告した先行研究は存在しない．
% JP: 構築したVRシステムが知覚的に許容可能な遅延範囲内で動作することを検証するため，本論文では視覚・触覚間遅延に関する実験結果も報告する．

\section{HaptoFlow}
\label{sec:model}

Following existing real-time vibrotactile generative models~\cite{transformer2022, DSTN2022, SPSI2024}, HaptoFlow takes three inputs at each rendering step: a material label, interaction parameters (2D stylus velocity and applied force), and the waveform generated at the previous timestep.
Given these inputs, the model outputs the vibrotactile waveform for the current timestep.
This section describes the architecture and training procedure of HaptoFlow.
% JP: 既存のリアルタイム振動触覚生成モデル~\cite{transformer2022, DSTN2022, SPSI2024}と同様に，HaptoFlowは各レンダリングステップにおいて3つの入力を受け取る：素材ラベル，インタラクションパラメータ（2Dスタイラス速度および押付力），そして前のタイムステップで生成された波形である。
% JP: これらの入力に基づき，モデルは現在のタイムステップの振動触覚波形を出力する。
% JP: 本節では，HaptoFlowのアーキテクチャと学習手順について述べる。

\subsection{Architecture}
\label{sec:model-arch}

The model consists of two main components: a Flow Matching module and a Signal Encoder-Decoder (\cref{fig:architecture} (a)).
% JP: \Cref{fig:architecture}にHaptoFlowのアーキテクチャの概要を示す。
% JP: 本モデルは，Flow Matchingモジュールと信号エンコーダ・デコーダの2つの主要コンポーネントから構成される。

The Flow Matching module is responsible for waveform generation.
Based on Flow Matching~\cite{FlowMatching2023}, it learns a conditional vector field in a latent space that transports a noise distribution toward the target waveform distribution.
This formulation enables efficient approximation of complex data distributions, allowing a compact model to achieve high waveform generation accuracy.
Here the noise sample acts as a random starting point for this transport. Because different noise samples under the same condition yield different plausible waveforms, the model represents the distribution of realistic haptic signals rather than collapsing to a single averaged output.
% JP: Flow Matchingモジュールは波形生成を担う。
% JP: 提供された入力を条件として，ノイズ分布を目標波形分布に輸送する潜在空間中のベクトル場を推定する。
% JP: Flow Matching~\cite{FlowMatching2023}は，連続時間の輸送過程を条件付きベクトル場として直接学習する生成モデリング手法である。
% JP: 推論時には，学習されたベクトル場をODEソルバにより数値的に積分することでサンプルが生成される。
% JP: この定式化により複雑なデータ分布の効率的な近似が可能となり，コンパクトなモデルで高い波形生成精度を達成できる。
% JP: ここでノイズサンプルはランダムな出発点として働き，同じ条件の下でも異なるノイズサンプルを輸送すれば異なる妥当な波形が得られるため，モデルは単一の平均化された出力に収束するのではなく，現実的な触覚信号の分布を表現できる。

Training is performed using the Conditional Flow Matching (CFM) loss~\cite{FlowMatching2023}.
Let $x_0 \sim \mathcal{N}(0, I)$ denote the noise sample, $x_1$ the target latent representation, and $c$ the conditioning information.
The interpolated sample at time $t \in [0,1]$ is defined as
\begin{equation}
  x_t = (1 - t)\,x_0 + t\,x_1.
\end{equation}
With the true conditional vector field $u_t(x_t \mid x_1) = x_1 - x_0$ and the network-predicted vector field $v_\theta$, the CFM loss is expressed as
\begin{equation}
  \mathcal{L}_{\mathrm{CFM}} = \mathbb{E}_{t,\, x_0,\, x_1}\left[\left\| v_\theta(x_t,\, t,\, c) - u_t(x_t \mid x_1) \right\|^2\right].
\end{equation}
% JP: 学習はConditional Flow Matching（CFM）損失~\cite{FlowMatching2023}を用いて行われる。
% JP: $x_0 \sim \mathcal{N}(0, I)$をノイズサンプル，$x_1$を目標潜在表現，$c$を条件情報とする。
% JP: 時刻$t \in [0,1]$における補間サンプルは上式で定義される。
% JP: 真の条件付きベクトル場$u_t(x_t \mid x_1) = x_1 - x_0$とネットワークが予測するベクトル場$v_\theta$を用いて，CFM損失は上式のように表される。

At inference, the learned field $v_\theta$ transports the noise sample $x_0$ toward the data distribution. The target latent $x_1$ is obtained by integrating the field from $t{=}0$ to $t{=}1$,
\begin{equation}
  x_1 = x_0 + \int_{0}^{1} v_\theta(x_t,\, t,\, c)\, \mathrm{d}t.
\end{equation}
An ODE solver approximates this integral over $N$ discrete steps, advancing from $x_0$ to $x_1$ by adding the field evaluated at each step.
% JP: 推論時には，学習済みベクトル場$v_\theta$がノイズサンプル$x_0$をデータ分布へと輸送する。目標潜在表現$x_1$は，ベクトル場を$t=0$から$t=1$まで積分することで上式のように得られる。
% JP: ODEソルバはこの積分を$N$個の離散ステップで近似し，各ステップで評価したベクトル場を加算しながら$x_0$から$x_1$へと進める。

We instantiate the Flow Matching module with a U-Net backbone~\cite{Unet2015}, a symmetric encoder-decoder architecture with skip connections widely used in flow-based generative models, that estimates the conditional vector field in the latent space (\cref{fig:architecture} (b)).
The U-Net has a two-level encoder-decoder structure with 128, 256 channels at each level, respectively.
To maintain waveform continuity between successive frames, the waveform generated at the previous frame is encoded into a latent representation by the Signal Encoder and concatenated with the feature map at each downsampling stage of the U-Net (\cref{fig:architecture}).
The conditioning information and a time embedding are each projected to 64-dimensional vectors: the material label via a learnable embedding layer, the interaction parameters via a multi-layer perceptron (MLP), and the Flow Matching time variable $t$ via a sinusoidal positional encoding.
These vectors are concatenated and injected into every level of the U-Net through Feature-wise Linear Modulation (FiLM)~\cite{FiLM2018}, an affine transformation that modulates intermediate features conditioned on external inputs.
At inference, the ODE solver is run for one step.
An ablation study examining the effect of the number of ODE-solver steps and the contribution of the Signal Encoder-Decoder is provided in the supplementary material.
% JP: Flow Matchingモジュールを，潜在空間における条件付きベクトル場を推定するU-Netバックボーンとして実装した。
% JP: ODEソルバのステップ数および信号エンコーダ・デコーダの寄与に関するアブレーションスタディは補足資料に示す。
% JP: U-Netは3段階のエンコーダ・デコーダ構造を持ち，各段階でそれぞれ128, 256, 256チャネルを有する。
% JP: 条件情報（素材ラベルおよびモーションパラメータ）とFlow Matchingの時間変数$t$の時間埋め込みは，それぞれ別の埋め込み層により64次元ベクトルに射影され，加算された後，Feature-wise Linear Modulation（FiLM）~\cite{FiLM2018}を通じてU-Netの全段階に注入される。
% JP: 推論時には，ODEソルバを1ステップで実行する。

The Signal Encoder-Decoder compresses haptic waveforms into a compact latent representation and reconstructs them from that representation.
By reducing the dimensionality of the input to the Flow Matching module, this component lowers both training cost and inference latency.
We use EnCodec~\cite{EnCodec2023} as the Signal Encoder-Decoder, an audio codec model that has previously been applied to haptic waveform generation~\cite{HapticGen2025}.
EnCodec is trained on diverse acoustic waveforms spanning speech, music, and environmental sounds, and achieves lightweight yet high-fidelity waveform compression and reconstruction through a convolutional encoder-decoder.
Our model uses the continuous latent representation from the encoder, prior to quantization, as the operating space for the Flow Matching module.
We integrate the pre-trained EnCodec model with weights frozen into our model, using the configuration with a 24\,kHz sampling rate and 6\,kHz bandwidth.
Since the frozen codec remains bound to the rate it was pre-trained on, we upsample our 2\,kHz haptic waveforms (\cref{sec:model-training}) to 24\,kHz before encoding and downsample the decoded output back.

% JP: 信号エンコーダ・デコーダは，触覚波形をコンパクトな潜在表現に圧縮し，その表現から波形を復元する。
% JP: Flow Matchingモジュールへの入力の次元を削減することにより，学習コストと推論遅延の両方を低減する。
% JP: 信号エンコーダ・デコーダとして，触覚波形生成に適用実績のある音声コーデックモデルEnCodec~\cite{EnCodec2023}を使用した~\cite{HapticGen2025}。
% JP: EnCodecは音声・音楽・環境音を含む多様な音響波形で学習されており，畳み込みエンコーダ・デコーダと残差ベクトル量子化の組み合わせにより，軽量かつ高忠実度の波形圧縮・復元を実現する。
% JP: 重みを凍結した事前学習済みEnCodecモデルを，24\,kHzサンプリングレートおよび6\,kHz帯域幅の設定で本モデルに統合した。
% JP: 凍結したコーデックは事前学習時のレートに固定されるため，2\,kHzの触覚波形（\cref{sec:model-training}）をエンコード前に24\,kHzへアップサンプリングし，デコード出力をダウンサンプリングして戻す。

\subsection{Training}
\label{sec:model-training}

\begin{figure}[t]
    \centering
    \includegraphics[width=1\linewidth]{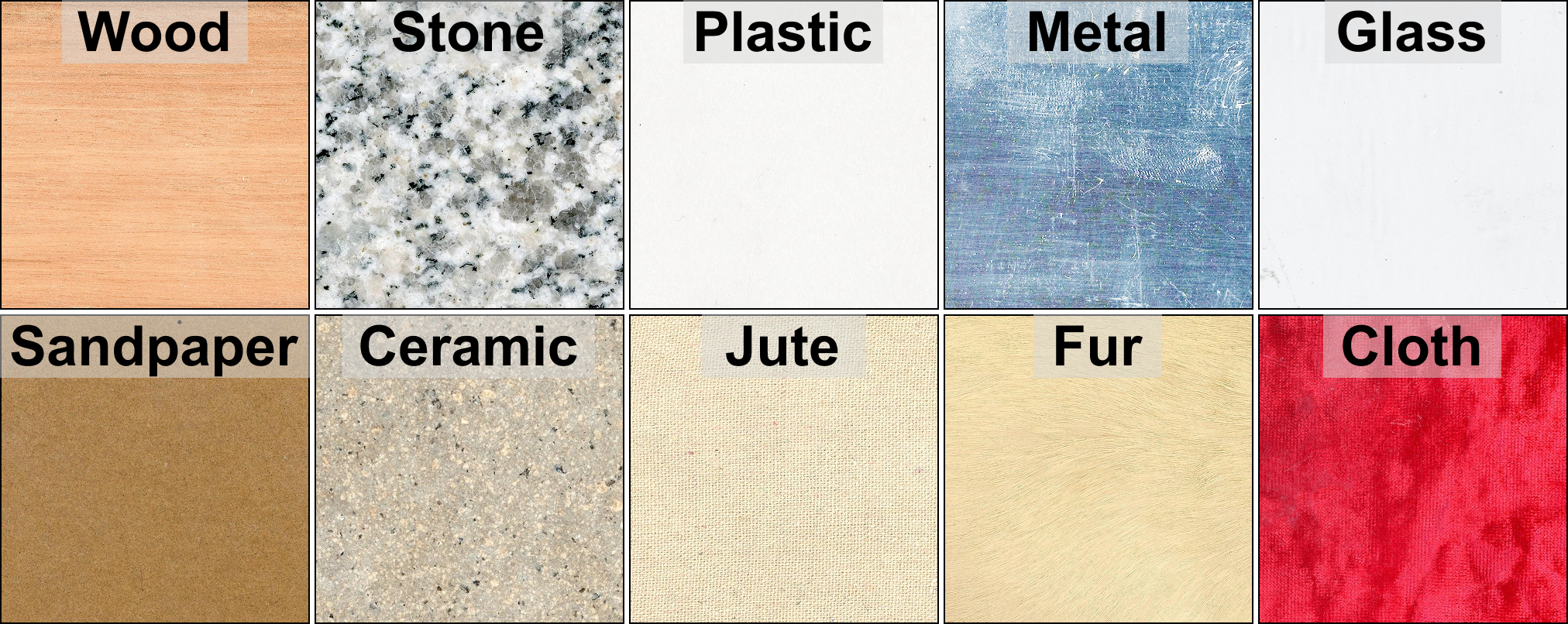}
    \caption{Materials used for training and evaluation.}
    % JP: 学習および評価に使用した素材。
    \label{fig:materials}
\end{figure}

We train the proposed model on the Cluster Haptic Texture Dataset~\cite{ClusterHapticTextureDataset2025}, which contains vibrotactile waveforms recorded from 118 distinct materials spanning 10 material categories, using a microphone and an accelerometer, along with corresponding interaction parameters (2D velocity vector and applied force).
One representative material was selected from each of the 10 categories to ensure diversity across distinct tactile characteristics (\cref{fig:materials}), and accelerometer recordings of these materials were used for training and evaluation.
The three-axis accelerometer data were reduced to a single axis using DFT321~\cite{DFT321}, a frequency-domain method that combines three-axis signals into a single perceptually representative axis.
% JP: 提案モデルをCluster Haptic Texture Dataset~\cite{ClusterHapticTextureDataset2025}で学習した。このデータセットは，マイクロフォンおよび加速度センサを用いて10の素材カテゴリにまたがる118種類の異なる素材から記録された振動触覚波形と，対応するインタラクションパラメータ（2D速度ベクトルおよび押付力）を含む。
% JP: 多様な触覚特性を確保するため，10カテゴリの各々から1つの代表的素材を選択し（\cref{fig:materials}），これらの素材の加速度センサ記録を学習および評価に使用した。
% JP: 3軸加速度データはDFT321~\cite{DFT321}を用いて1軸に圧縮した。

% Fig: 素材の表面画像

During preprocessing, each waveform was first downsampled to 2\,kHz and then segmented into 100\,ms frames.
To increase the volume of training data, a sliding window with varying start positions was applied during segmentation.
Waveform amplitudes were normalized to the range $[-1, 1]$ using the maximum amplitude observed across all waveforms in the dataset as the normalization reference.
% JP: 前処理では，各波形をまず2\,kHzにダウンサンプリングし，100\,msのフレームに分割した。
% JP: 学習データの量を増やすため，分割時に開始位置を変化させるスライディングウィンドウを適用した。
% JP: 波形振幅は，データセット内の全波形において観測された最大振幅を正規化基準として，$[-1, 1]$の範囲に正規化した。

\section{VR System with Vibrotactile Generative Model}
To demonstrate the proposed vibrotactile generative model in an interactive setting, we integrated it into a VR system consisting of a head-mounted display (HMD) and a haptic-enabled stylus.
This section describes the system architecture and characterizes the worst-case visual-haptics delay of the complete pipeline.
% JP: 提案する触覚生成モデルをインタラクティブな環境で実証するため，ヘッドマウントディスプレイ（HMD）と触覚対応スタイラスから構成されるVRシステムに統合した。
% JP: 本節では，システムアーキテクチャについて述べるとともに，パイプライン全体の最悪の視覚・触覚間遅延を特性評価する。

\subsection{System Overview}
\begin{figure}[t]
  \centering
  \includegraphics[width=1\linewidth]{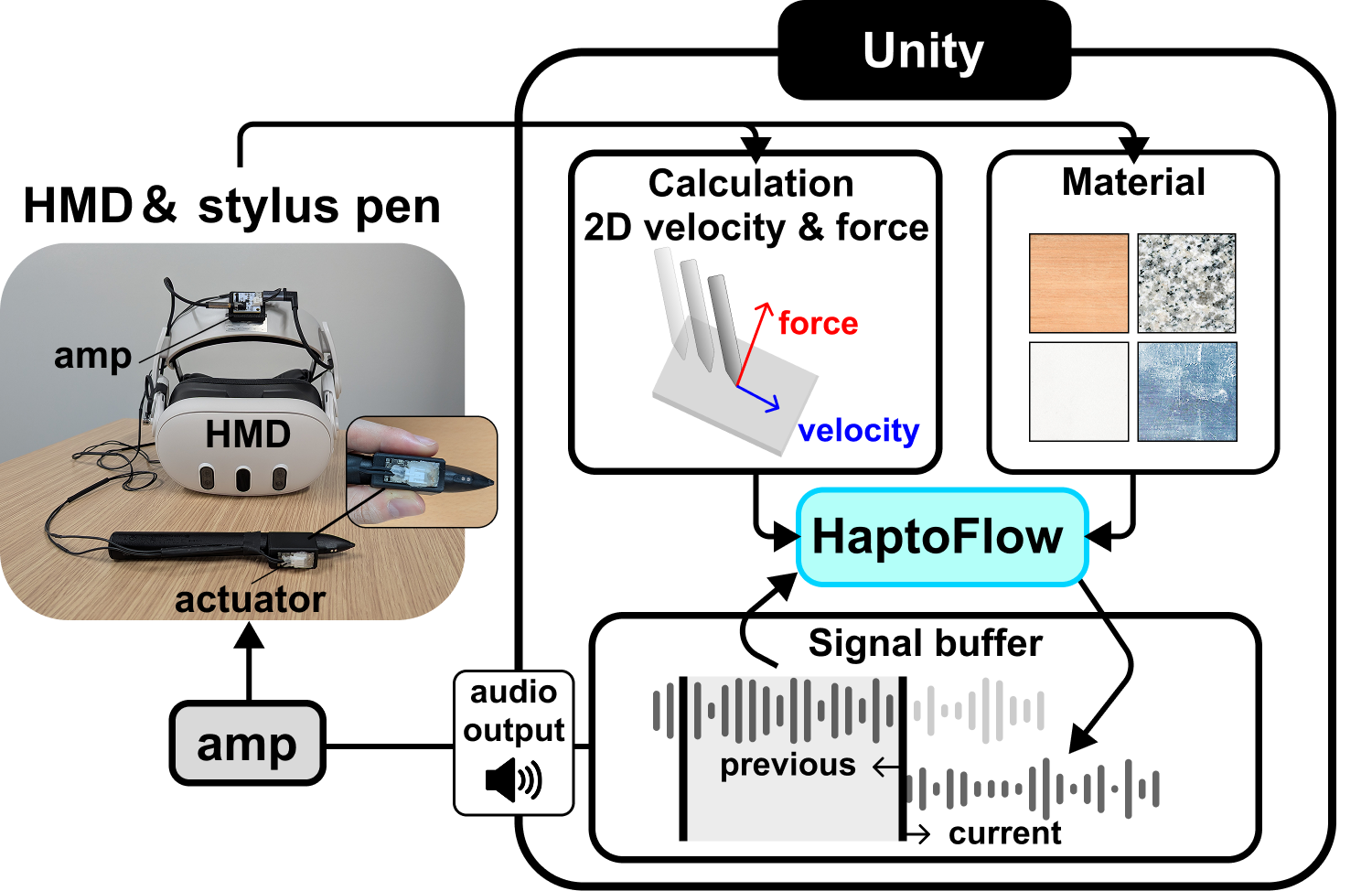}
  \caption{Overview of the VR haptic system. At each rendering step, HaptoFlow receives stylus velocity, contact force, a material label, and the previously generated waveform to produce real-time vibrotactile feedback via an actuator-equipped stylus.}
  % JP: VR触覚システムの概要。各レンダリングステップにおいて，HaptoFlowはスタイラス速度，接触力，素材ラベル，および前に生成された波形を受け取り，アクチュエータ搭載スタイラスを介してリアルタイムの振動触覚フィードバックを生成する。
  \label{fig:VR_system}
\end{figure}

\Cref{fig:VR_system} shows an overview of the system.
The VR environment is rendered on a Meta Quest~3 HMD.
For stylus input, we use the MX Ink (Logicool), which provides 3D positional tracking when paired with Meta Quest~3.
A vibrotactile actuator (Haptic Reactor, Foster Electric) is embedded in the stylus body and serves as the haptic output device.
The output waveform produced by the vibrotactile generative model is amplified by a PAM8012 amplifier (Diodes Incorporated) before being delivered to the actuator.
% JP: \Cref{fig:VR_system}にシステムの概要を示す。
% JP: VR環境はMeta Quest~3 HMD上にレンダリングされる。
% JP: スタイラス入力にはMX Ink（Logicool）を使用し，Meta Quest~3とペアリングすることで3D位置トラッキングを提供する。
% JP: 振動触覚アクチュエータ（Haptic Reactor, Foster Electric）がスタイラス本体に埋め込まれ，触覚出力デバイスとして機能する。
% JP: 触覚生成モデルが出力した波形は，PAM8012アンプ（Diodes Incorporated）で増幅された後，アクチュエータに送られる。

When the stylus contacts a virtual object, the system computes two signals that serve as inputs to the vibrotactile generative model.
The first is the 2D stylus velocity relative to the object surface, obtained by projecting the 3D stylus velocity onto the tangent plane of the contact surface using the surface normal at the contact point.
The second is the applied force, approximated using a spring model ($f = k \cdot d$, where $k$ is the spring stiffness and $d$ is the penetration depth of the stylus into the virtual surface).
These two signals, together with the material label and the previous waveform, are passed to the vibrotactile generative model at each rendering step.
The generated waveform, 100\,ms in length, is stored in a buffer, from which the previous waveform input is extracted at each update according to the system's rendering interval; only the leading segment matching that interval is output before the next generation overwrites the remainder.
At the initial step, when the buffer is empty, a waveform with matching generation conditions is retrieved from the training dataset and used as the previous waveform input, following the approach of \cite{DSTN2022}.
The output waveform is then routed to the actuator via Unity's audio output system, enabling synchronous vibrotactile feedback during stylus-surface interaction.
% JP: スタイラスが仮想オブジェクトに接触すると，システムは触覚生成モデルへの入力となる2つの信号を計算する。
% JP: 第一はオブジェクト表面に対する2Dスタイラス速度であり，接触点における法線を用いて3Dスタイラス速度を接触面の接平面に射影することで得られる。
% JP: 第二は押付力であり，ばねモデル（$f = k \cdot d$，$k$はばね剛性，$d$はスタイラスの仮想表面への貫入深さ）を用いて近似する。
% JP: これら2つの信号は，素材ラベルおよび前に生成された波形フレームとともに，各レンダリングステップで触覚生成モデルに渡される。
% JP: 生成された波形（長さ100\,ms）はバッファに格納され，そこからシステムのレンダリング間隔に応じて各更新時のprevious waveform入力が取り出される。実際に出力されるのは同間隔に相当する先頭部分のみであり，残りは次の生成により上書きされる。
% JP: 出力波形はUnityのオーディオ出力システムを介してアクチュエータに送られ，スタイラスと表面のインタラクション中に同期した振動触覚フィードバックを実現する。

\subsection{Worst-case Latency of Haptic Device}
\label{sec:worst-case-latency}

We characterize the visual-haptics delay of our system by measuring two fixed hardware delay components.
The first component is the controller pose update interval, which determines how quickly a contact event detected visually can be forwarded to the haptic generation pipeline.
Meta Quest~3 supports refresh rates from 72\,Hz to 120\,Hz; we configured it to 72\,Hz to accommodate the computational load of the haptic generative module, yielding a maximum pose update interval of 14\,ms.
The second component is the mechanical response time of the haptic actuator.
We measured both the rise time (signal onset to maximum amplitude) and the fall time (signal termination to full stop) by attaching a triaxial accelerometer (ADXL335, Analog Devices) to the actuator housing and recording the output on a digital oscilloscope (MSO5074, RIGOL).
Both rise and fall times were approximately 3\,ms.
These two hardware-bound delays sum to 17\,ms regardless of the haptic generation algorithm employed.
Adding a vibrotactile generative model with inference latency $X$\,ms (reported in \cref{tab:results}) therefore yields a worst-case end-to-end visual-haptics delay of $(17 + X)$\,ms.
% JP: 本システムの視覚・触覚間遅延を，2つの固定的なハードウェア遅延成分を測定することで特性評価した。
% JP: 第一の成分はコントローラの姿勢更新間隔であり，Meta Quest~3を72\,Hzに設定し最大14\,msとなった。
% JP: 第二の成分は触覚アクチュエータの機械的応答時間であり，約3\,msであった。
% JP: これら2つのハードウェア遅延は合計17\,msとなり，推論遅延$X$\,ms（\cref{tab:results}に報告）の生成モデルを加えると，最悪のエンドツーエンド視覚・触覚間遅延は$(17 + X)$\,msとなる。

\section{Technical Evaluation}
% 評価概要（目的、実施内容（モデルの出力波形の再現精度、推論時間））
We conducted a comparative study against existing vibrotactile generative models to evaluate the effectiveness of the proposed model, assessing waveform reproduction accuracy and inference latency.
This experiment tests the following hypothesis:

\noindent\textbf{H1}: \textit{The proposed Flow Matching-based model achieves (a) higher waveform reproduction accuracy (GFC, RMSE) and (b) lower inference latency than the baseline methods trained with deterministic reconstruction losses.}
% JP: 本実験は以下の仮説を検証する：
% JP: \textbf{H1}: 提案するFlow Matchingベースのモデルは，決定論的再構成損失で学習されたベースライン手法と比較して，(a) より高い波形再現精度（GFC, RMSE）と (b) より低い推論遅延を同時に達成する．
% JP: 提案モデルの有効性を評価するため、既存の振動触覚生成モデルとの比較研究を実施し、
% JP: 波形再現精度および推論レイテンシを評価した。

\subsection{Evaluation Setup}
\subsubsection{Experimental Conditions}
We used the Cluster Haptic Texture Dataset~\cite{ClusterHapticTextureDataset2025}, partitioned into training, validation, and evaluation subsets at a 70:10:20 ratio, to train and evaluate each vibrotactile generative model.
Training was conducted on a machine equipped with an Intel Xeon Gold 6326 CPU, an NVIDIA A100 (80\,GB) GPU, and 512\,GB of RAM.
Inference was evaluated on a separate machine with an Intel Core Ultra 9 285 CPU, an NVIDIA GeForce RTX 5090 GPU, and 64\,GB of RAM.
Both machines ran Ubuntu 22.04.
Each model was trained and evaluated five times with different random seeds; reproduction accuracy and inference latency are reported as the mean and standard deviation across these runs.
All models were optimized using AdamW~\cite{AdamW2019} with a learning rate of $10^{-4}$, a batch size of 128, and trained for 30 epochs.
% JP: Cluster Haptic Texture Dataset~\cite{ClusterHapticTextureDataset2025}を使用し、訓練・検証・評価サブセットを70:10:20の比率で分割して各触覚生成モデルの訓練および評価を行った。
% JP: 訓練はIntel Xeon Gold 6326 CPU、NVIDIA A100（80\,GB）GPU、512\,GBのRAMを搭載したマシンで実施した。
% JP: 推論はIntel Core Ultra 9 285 CPU、NVIDIA GeForce RTX 5090 GPU、64\,GBのRAMを搭載した別のマシンで評価した。
% JP: 両マシンともUbuntu 22.04を使用した。
% JP: 各モデルは異なるランダムシードで5回訓練・評価を行い、再現精度および推論レイテンシはこれらの実行の平均と標準偏差として報告する。
% JP: すべてのモデルはAdamW~\cite{AdamW2019}を用い、学習率$10^{-4}$、バッチサイズ128で30エポック訓練した。

\subsubsection{Evaluation Metrics}
We evaluated each model using three metrics: the Goodness-of-Fit Criterion (GFC)~\cite{abdulali2016}, which measures waveform reproduction accuracy in the frequency domain; the Root Mean Square Error (RMSE), which measures reproduction accuracy in the time domain; and inference latency (ms).
% JP: 各モデルを3つの指標で評価した。周波数領域における波形再現精度を測定するGoodness-of-Fit Criterion（GFC）~\cite{abdulali2016}、
% JP: 時間領域における再現精度を測定するRoot Mean Square Error（RMSE）、および推論レイテンシ（ms）である。

GFC is defined as the normalized inner product of the amplitude spectra of the reference and generated waveforms~\cite{abdulali2016}, ranging from 0 to 1, with values closer to 1 indicating greater spectral agreement.
% JP: GFCは参照波形と生成波形の振幅スペクトルの正規化内積として定義され~\cite{abdulali2016}，0から1の範囲をとり，1に近いほどスペクトルの一致度が高い。

RMSE is the square root of the mean squared error between the reference waveform $d(t)$ and the generated waveform $m(t)$ in the time domain; lower values indicate closer agreement with the reference waveform.
Inference latency was defined as the elapsed time from receiving the model inputs to obtaining the output waveform.
% JP: RMSEは時間領域における参照波形$d(t)$と生成波形$m(t)$の平均二乗誤差の平方根であり、値が低いほど参照波形との一致度が高いことを示す。
% JP: 推論レイテンシは、モデルへの入力を受け取ってから出力波形を得るまでの経過時間として定義した。

\subsubsection{Baseline Methods}
We compared the proposed model against three baselines, each trained on the same dataset with hyperparameters matching the respective original papers:
\textbf{Transformer}~\cite{transformer2022}, a Transformer Encoder-Decoder that captures long-range temporal dependencies via self-attention;
\textbf{DSTN}~\cite{DSTN2022}, a 1D CNN combined with an LSTM Encoder-Decoder for joint spectral-temporal modeling;
and \textbf{SPSI}~\cite{SPSI2024}, an MLP-based amplitude spectrum predictor with non-iterative SPSI phase recovery.
% JP: 提案モデルを以下の3手法と比較した。各手法は提案モデルと同一のデータセットで訓練・評価し、
% JP: ハイパーパラメータはそれぞれの原論文で報告された値に合わせて設定した。
% JP: \textbf{Transformer}~\cite{transformer2022}：自己注意機構によるEncoder-Decoderアーキテクチャを用いて長距離の時間的依存性を捕捉する。
% JP: \textbf{DSTN (Deep Spatial-Temporal Network)}~\cite{DSTN2022}：触覚波形の局所的な周波数特徴を抽出する1D CNNと、
% JP: 大局的な時間ダイナミクスをモデル化するLSTMベースのEncoder-Decoderを組み合わせ、スペクトル特性と時系列変動の同時モデル化を可能にする。
% JP: \textbf{SPSI (Single Pass Spectrogram Inversion)}~\cite{SPSI2024}：MLPを用いて離散フーリエ変換の振幅スペクトルを予測し、
% JP: 非反復的なSPSI位相復元アルゴリズムにより時間領域波形を再構成する。
Transformer and SPSI identify the material from images (a texture image~\cite{transformer2022} and GelSight input~\cite{SPSI2024}), which would conflate haptic generation with image-based material recognition, so we replaced both with a uniform one-hot material label and added the same input to DSTN~\cite{DSTN2022}, which originally accepts no material label.
Baselines are otherwise as published: Transformer and DSTN take the previous-timestep waveform, whereas SPSI omits it to minimize latency, a spread of design choices we retain for comparison.
% JP: TransformerとSPSIは素材を画像から識別しており（それぞれテクスチャ画像~\cite{transformer2022}とGelSight入力~\cite{SPSI2024}），これは触覚生成と画像ベースの素材認識を交絡させるため，両者を統一的なone-hot素材ラベルに置き換え，元々素材ラベル入力を持たないDSTN~\cite{DSTN2022}にも同じ入力を追加した．
% JP: それ以外は各ベースラインを原論文のまま用いた：TransformerとDSTNはprevious-timestep波形を入力に取るが，SPSIはレイテンシ最小化のためこれを省いており，この設計選択の幅を比較のために保持している．

\subsection{Result}
% table: 各手法の再現精度、推論時間
\begin{table}[t]
  \centering
  \caption{Waveform reproduction accuracy and inference latency of each vibrotactile generative model}
  % JP: 各振動触覚生成モデルの波形再現精度および推論レイテンシ
  \label{tab:results}
  \begin{tabular}{@{}lccc@{}}
    \toprule
    Model & GFC ↑ & RMSE ↓ & Latency (ms) ↓ \\
    \midrule
    \textbf{Ours}      & \textbf{0.96 $\pm$ 0.04} & \textbf{0.22 $\pm$ 0.33} & \textbf{5.2 $\pm$ 1.0} \\
    Transformer~\cite{transformer2022}        & 0.74 $\pm$ 0.17          & 0.39 $\pm$ 0.41          & 13.4 $\pm$ 3.4          \\
    DSTN~\cite{DSTN2022} & 0.94 $\pm$ 0.05          & 0.30 $\pm$ 0.36         & 11.9 $\pm$ 2.8          \\
    SPSI~\cite{SPSI2024} & 0.73 $\pm$ 0.15 & 0.49 $\pm$ 0.44          & 6.2 $\pm$ 1.8         \\
    \bottomrule
  \end{tabular}
\end{table}

% fig: 各手法の出力波形
\begin{figure*}[!t]
    \centering
    \includegraphics[width=1\linewidth]{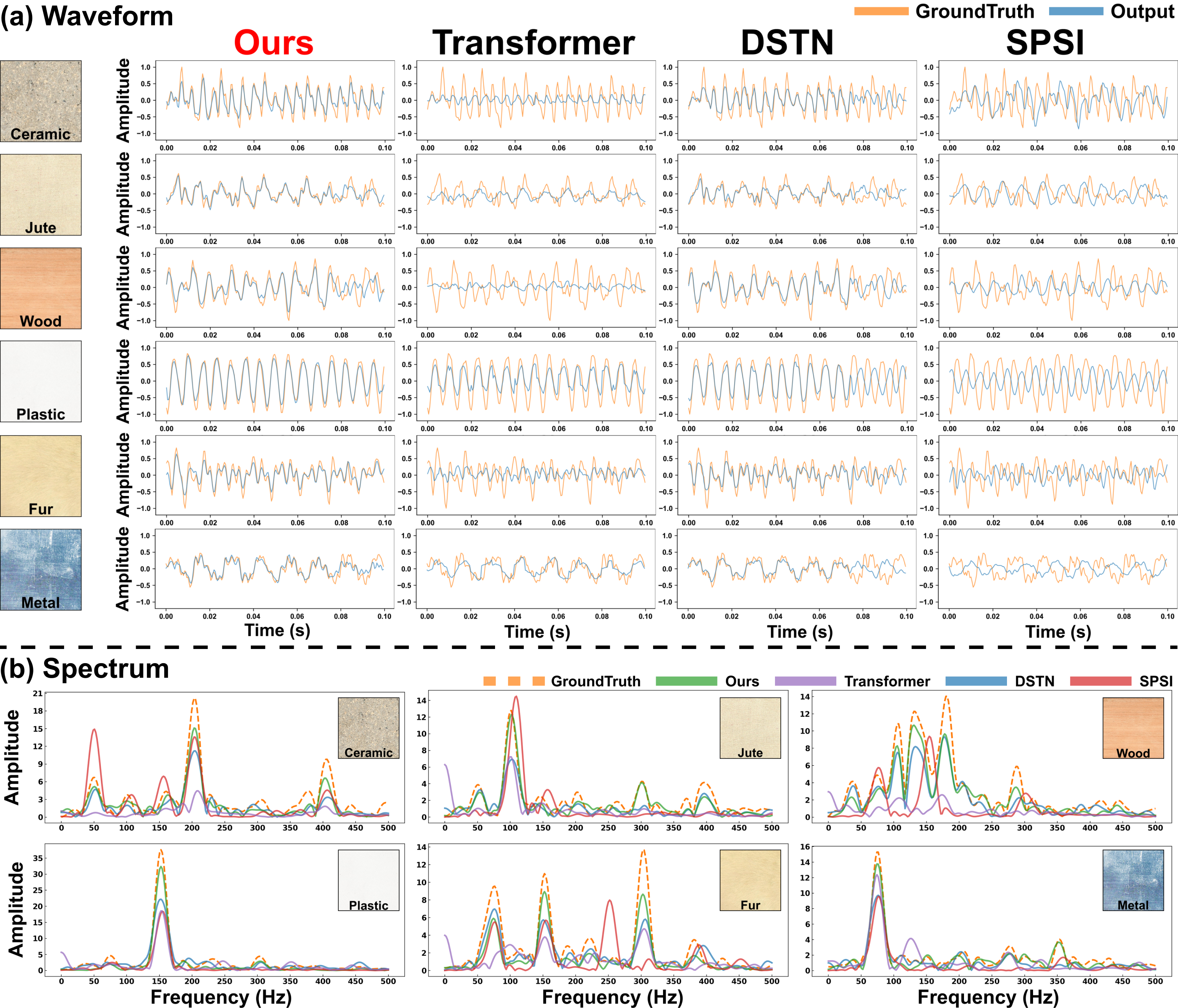}
    \caption{Ground truth and output waveforms of each model for six representative samples, shown as (a) raw time-domain signals and (b) frequency spectra zoomed into the 0--500\,Hz range relevant to vibrotactile perception.}
    % JP: 6つの代表的サンプルに対する各モデルの正解波形と出力波形。(a)生の時間領域信号、(b)振動触覚知覚に関連する0--500\,Hz範囲に拡大した周波数スペクトルを示す。
    \label{fig:results}
\end{figure*}

\Cref{tab:results} summarizes the evaluation results for all models.
The proposed model achieves the highest GFC and lowest RMSE among all methods, demonstrating superior waveform reproduction accuracy.
It also achieves the shortest inference latency at 5.2\,ms, outperforming all baselines in responsiveness.
The output waveforms shown in \cref{fig:results} further confirm that the proposed model most faithfully reproduces fine temporal variations of the reference signal.
These results demonstrate that the proposed Flow Matching-based model achieves superiority in both waveform reproduction accuracy and inference latency.
% JP: \Cref{tab:results}に全モデルの評価結果をまとめる。
% JP: 提案モデルは全手法中最高のGFCおよびRMSEを達成し、優れた波形再現精度を実証した。
% JP: また、推論レイテンシも5.2\,msと最短であり、応答性においても全ベースラインを上回った。
% JP: \cref{fig:results}に示す出力波形は、提案モデルが参照信号の微細な時間変動を最も忠実に再現していることをさらに確認するものである。
% JP: これらの結果は、提案するFlow Matchingベースのモデルが波形再現精度と推論レイテンシの両方において優位性を達成することを実証している。
These results support H1; a detailed discussion is provided in \cref{sec:discussion}.
% JP: これらの結果はH1を支持する。詳細な議論は\cref{sec:discussion}にて行う。

\section{User Study}
We conducted two user studies. The first investigated the perceptual threshold for visual-haptic delay (\cref{sec:study-latency}), testing:

\noindent\textbf{H2}: \textit{The total end-to-end system latency, including the inference time of the proposed model, falls below the human perceptual threshold for visual-haptic delay when using a VR HMD with a stylus pen.}
% JP: \textbf{H2}: 提案モデルの推論時間を含むシステム全体のエンドツーエンド遅延は，VR HMDとスタイラスペンの組み合わせにおけるヒトの視覚-触覚遅延知覚閾値を下回る．

The second assessed perceptual similarity between generated and real haptic stimuli (\cref{sec:study-similarity}), testing:

\noindent\textbf{H3}: \textit{Haptic stimuli generated by the proposed model receive higher perceptual similarity ratings to real objects than those generated by baseline methods.}
% JP: \textbf{H3}: 提案モデルにより生成された触覚刺激は，ベースライン手法により生成された触覚刺激と比較して，実物体に対するより高い知覚的類似度評価を得る．

% JP: 提案する振動触覚生成モデルを評価するため、2つのユーザスタディを実施した。
% JP: 第1の実験では、振動触覚生成モデルの推論時間を含むエンドツーエンドのシステムレイテンシの合計が、VR環境においてスタイラスペンを使用した際の視覚-触覚遅延に対する人間の知覚閾値を下回るかどうかを調査した（\cref{sec:study-latency}）。
% JP: 第2の実験では、提案手法およびベースライン手法により生成された触覚刺激と実物体から生じる触覚刺激との知覚的類似度を評価した（\cref{sec:study-similarity}）。

Both studies were approved by the Cluster, Inc. Research Ethics Committee (Registration number: 2025-012).
% JP: 両実験は著者らの所属機関の倫理審査委員会の承認を得て実施した（承認番号：[査読のため匿名化]）。

\subsection{User Study on Latency Perception}
\label{sec:study-latency}

In this study, we asked participants to report whether they perceived a delay between their stylus pen motion and the onset or cessation of haptic feedback under varying levels of artificially introduced latency.
% JP: 本実験では、人為的に導入されたさまざまなレベルのレイテンシの下で、スタイラスペンの動きと触覚フィードバックの開始または停止の間に遅延を知覚したか否かを参加者に報告させた。

\begin{figure}[t]
  \centering
  \includegraphics[width=1\linewidth]{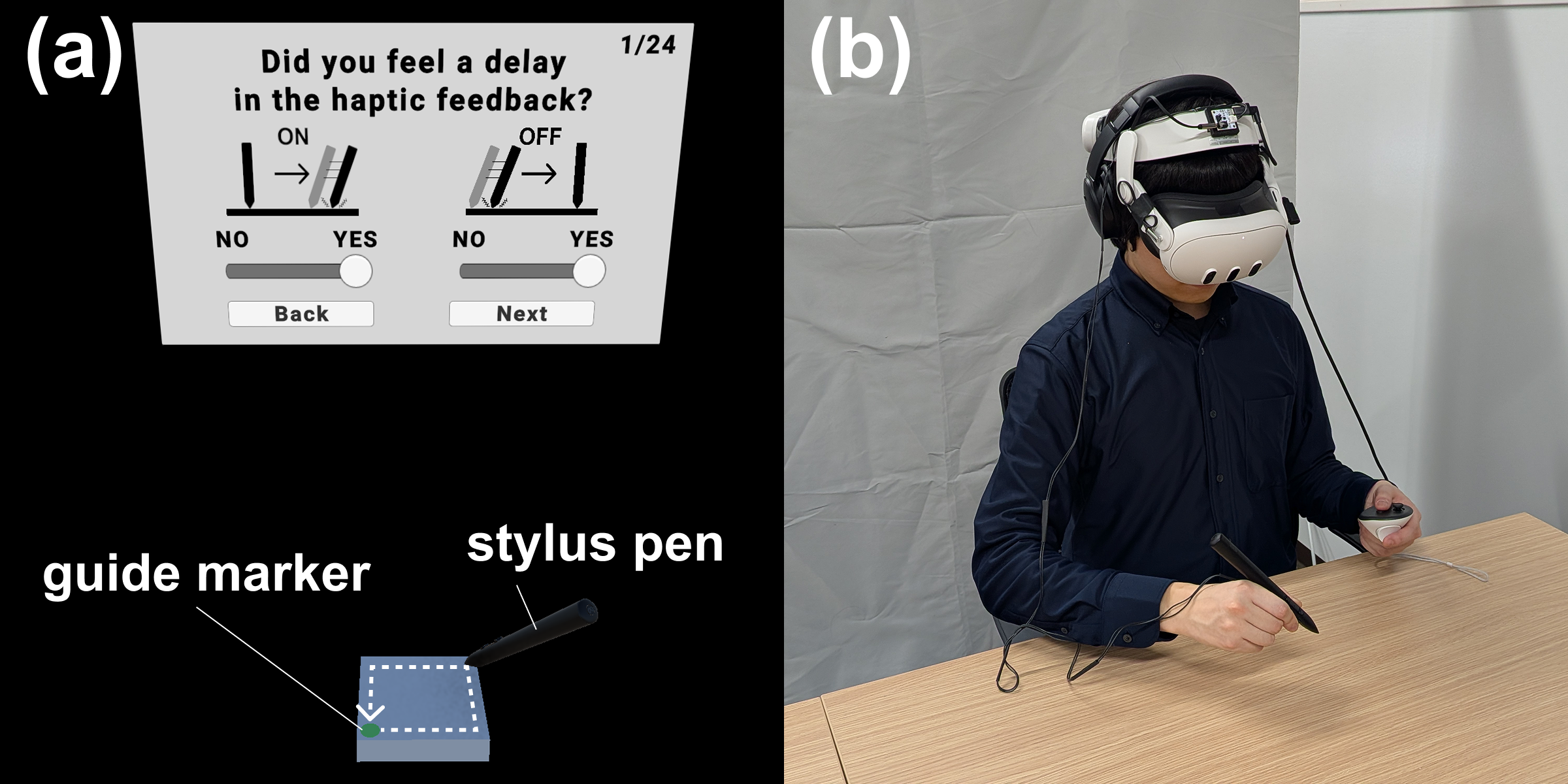}
  \caption{Experimental situation of the user study on latency perception. (a) VR scene viewed through the HMD, with a virtual object and questionnaire form on a dark background. (b) A participant during the experiment.}
  % JP: レイテンシ知覚ユーザスタディの実験状況。(a) HMDを通して見たVRシーン。暗い背景上に仮想物体と質問票フォームが表示されている。(b) 実験中の参加者。
  \label{fig:latensy_setup}
\end{figure}

\subsubsection{Apparatus and Task}

The experiment was conducted in a PC VR environment, in which a Meta Quest 3 HMD was connected to a desktop PC via a Meta Quest Link cable.
As shown in \cref{fig:latensy_setup}, each participant wore a VR HMD and noise-canceling headphones playing white noise to mask any auditory cues from the haptic device.
They held a stylus pen in their right hand for tracing a virtual object rendered in VR space, and a VR controller in their left hand for responding to on-screen questionnaires.
A green marker moved across the object surface at 150\,mm/s along a predefined trajectory, and participants were instructed to trace the marker with the stylus pen with a slight lag.
As participants traced the object surface, they received haptic feedback from a vibrator attached to the stylus pen.
The haptic stimulus was a sinusoidal waveform at 200\,Hz with a fixed amplitude, delivered whenever the pen was in motion on the object surface, regardless of pen speed or applied force.
% JP: \cref{fig:latensy_setup} に示すように、各参加者はVR HMDおよびハプティックデバイスからの聴覚的手がかりを遮蔽するためにホワイトノイズを再生するノイズキャンセリングヘッドホンを装着した。
% JP: 右手にはVR空間に描画された仮想物体をなぞるためのスタイラスペンを、左手には画面上の質問票に回答するためのVRコントローラを持った。
% JP: 緑色のマーカーがあらかじめ定義された軌跡に沿って150\,mm/sで物体表面上を移動し、参加者はわずかな遅れをもってスタイラスペンでマーカーをなぞるよう教示された。
% JP: 参加者が物体表面をなぞると、スタイラスペンに取り付けられたバイブレータから触覚フィードバックが提示された。
% JP: 触覚刺激は200\,Hz、固定振幅の正弦波であり、ペンの速度や押付力にかかわらず、ペンが物体表面上で運動している間は常に提示された。

The artificially introduced delay ranged from 50\,ms to 250\,ms in 20\,ms increments, yielding 11 delay levels.
All delay values represent total end-to-end system latency, including the inference time of the vibrotactile generative model.
We tested two conditions based on the haptic state change: \textit{Turn-On}, in which the delay was measured from when the pen began moving from a stationary state until the participant perceived the onset of haptic feedback; and \textit{Turn-Off}, in which the delay was measured from when the pen came to a stop until the participant perceived the cessation of haptic feedback.
% JP: 人為的に導入された遅延は50\,msから250\,msまで20\,ms刻みで設定し、11段階の遅延レベルを設けた。
% JP: すべての遅延値は、触覚生成モデルの推論時間を含むエンドツーエンドのシステムレイテンシの合計を表す。
% JP: 触覚状態の変化に基づき2つの条件を設定した：\textit{Turn-On} はペンが静止状態から動き始めてから参加者が触覚フィードバックの開始を知覚するまでの遅延を計測する条件、\textit{Turn-Off} はペンが停止してから参加者が触覚フィードバックの停止を知覚するまでの遅延を計測する条件である。

\subsubsection{Participants and Procedure}

A priori power analysis (G*Power~3.1) for the Friedman and Wilcoxon signed-rank tests planned in the perceptual similarity study (\cref{sec:study-similarity}), which shares the same participant pool, indicated a required sample of 24 ($f = 0.25$, $d_z = 0.67$, $\alpha = 0.05$, power $= 0.80$).
% JP: 知覚的類似度実験（\cref{sec:study-similarity}，同一参加者プール）で計画したFriedman検定・Wilcoxon符号順位検定に対するG*Power~3.1による事前検定力分析の結果，必要サンプルサイズは24名であった（$f = 0.25$，$d_z = 0.67$，$\alpha = 0.05$，検定力$= 0.80$）．

Twenty-four participants (14 male, 10 female; age range: 18--26 years; mean age: 22.3 years, SD: 2.1) took part in this study.
Eighteen of them had prior VR experience, and all participants were right-handed with normal or corrected vision.
% JP: 24名の参加者（男性14名、女性10名；年齢範囲：18--26歳；平均年齢：22.3歳、SD：2.1）が本実験に参加した。
% JP: うち18名はVR経験を有しており、全参加者が右利きで正常または矯正視力を有していた。

Before the experiment, participants received instructions on how to operate the VR system, followed by a practice session to familiarize themselves with the task.
Participants responded to a total of 22 conditions (11 delay levels $\times$ 2 conditions: Turn-On and Turn-Off) by indicating for each whether they perceived a delay (Yes/No).
The order of delay levels was counterbalanced and randomized across participants.
% JP: 実験前に、参加者はVRシステムの操作方法に関する説明を受け、続いてタスクに慣れるための練習セッションを行った。
% JP: 参加者は合計22条件（11遅延レベル $\times$ 2条件：Turn-On および Turn-Off）に対し、それぞれ遅延を知覚したか否か（Yes/No）を回答した。
% JP: 遅延レベルの提示順序は参加者間でカウンターバランスおよびランダム化した。

\subsubsection{Results}
\begin{figure}[t]
    \centering
    \includegraphics[width=1\linewidth]{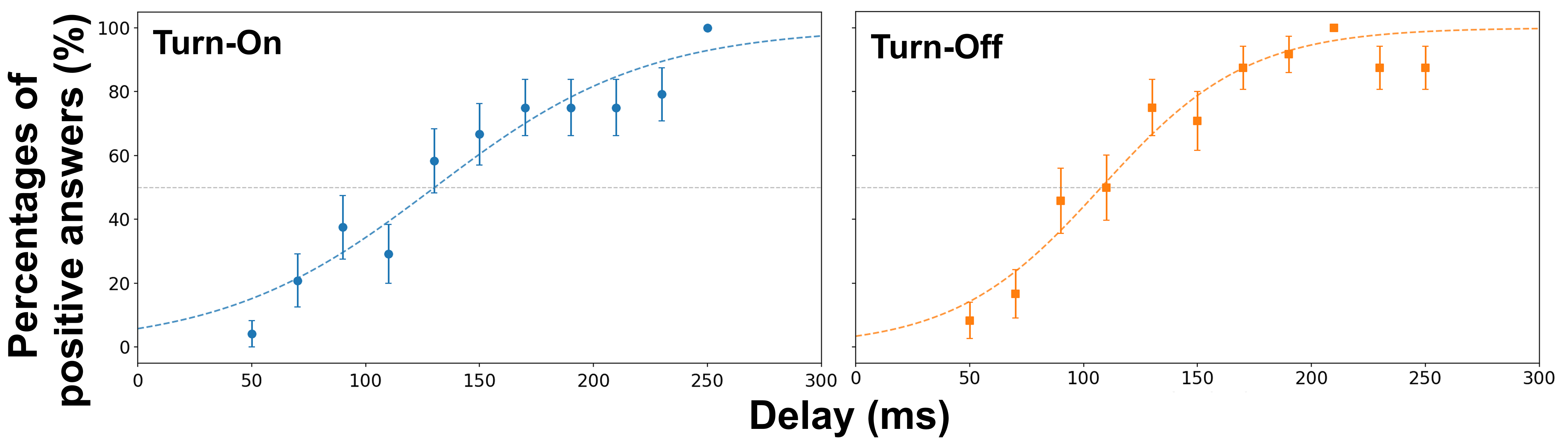}
    \caption{Percentages of positive answers for latency perception under the \textit{Turn-On} and \textit{Turn-Off} conditions. Each point represents the mean and error bars indicate the standard error. The dashed curves are fitted psychometric functions.}
    % JP: \textit{Turn-On} および \textit{Turn-Off} 条件におけるレイテンシ知覚の肯定回答率。各点は平均値を、誤差棒は標準誤差を示す。破線はフィッティングされた心理測定関数である。
    \label{fig:latensy}
\end{figure}

\Cref{fig:latensy} shows the proportion of participants who perceived a delay at each delay level for the Turn-On and Turn-Off conditions.
Following \cite{HaptoFloater2024}, we fitted a sigmoid function $y = 100 / (1 + \exp(-k(x - x_0)))$ to the data, where $x$ is the introduced delay (ms) and $y$ is the percentage of participants reporting a perceived delay.
% JP: \cite{HaptoFloater2024}に従い，シグモイド関数$y = 100 / (1 + \exp(-k(x - x_0)))$をデータにフィッティングした．$x$は導入遅延（ms），$y$は遅延を知覚した参加者の割合（\%）である．

The fitting parameters were $k = 0.0214$, $x_0 = 130.25$\,ms for the Turn-On condition, and $k = 0.0311$, $x_0 = 107.74$\,ms for the Turn-Off condition.
Following \cite{HaptoFloater2024}, we defined the perceptual delay threshold as the delay level at which the fitted sigmoid exceeds 50\%, yielding thresholds of 130.3\,ms and 107.7\,ms for the Turn-On and Turn-Off conditions, respectively.
% JP: フィッティングパラメータは、Turn-On 条件で $k = 0.0214$, $x_0 = 130.25$\,ms、Turn-Off 条件で $k = 0.0311$, $x_0 = 107.74$\,ms であった。
% JP: \cite{HaptoFloater2024} に従い、フィッティングされたシグモイド関数が50\%を超える遅延レベルを知覚遅延閾値と定義した結果、Turn-On 条件で130.3\,ms、Turn-Off 条件で107.7\,ms の閾値が得られた。

These results are discussed in relation to H2 in \cref{sec:discussion}.
% JP: これらの結果はH2との関連で\cref{sec:discussion}にて議論する。

\begin{figure}[t]
    \centering
    \includegraphics[width=1\linewidth]{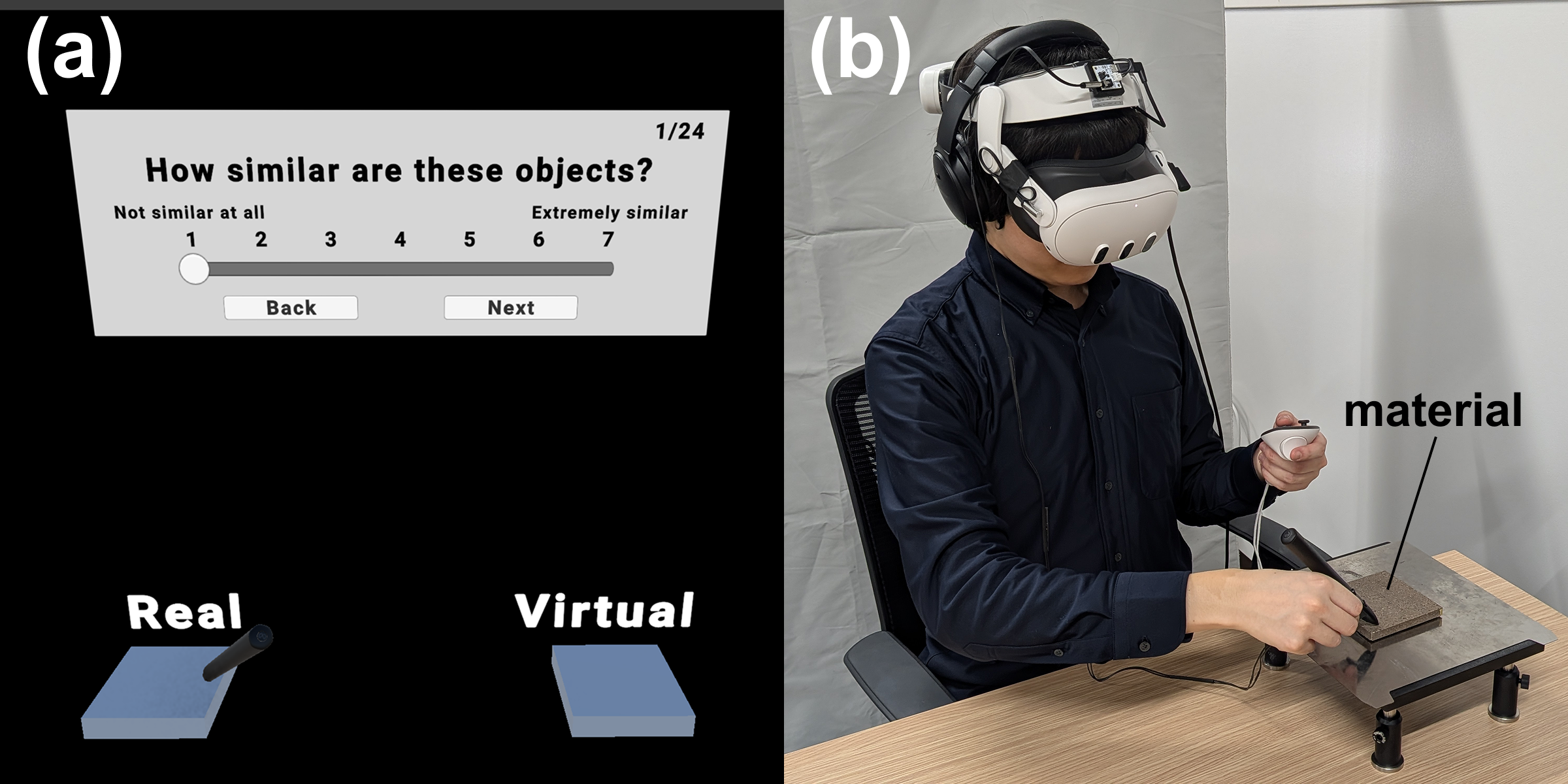}
    \caption{Experimental situation of the perceptual similarity user study. (a) VR scene showing two visually identical objects (real and virtual) and a similarity rating form. (b) A participant stroking a physical material while wearing the HMD, without seeing its appearance.}
    % JP: 知覚的類似度ユーザスタディの実験状況。(a) 視覚的に同一の2つの物体（実物と仮想）および類似度評価フォームを表示したVRシーン。(b) HMDを装着し、外観を見ずに物理素材をなぞる参加者。
    \label{fig:similarity_setup}
\end{figure}
\begin{figure*}[t]
    \centering
    \includegraphics[width=1\linewidth]{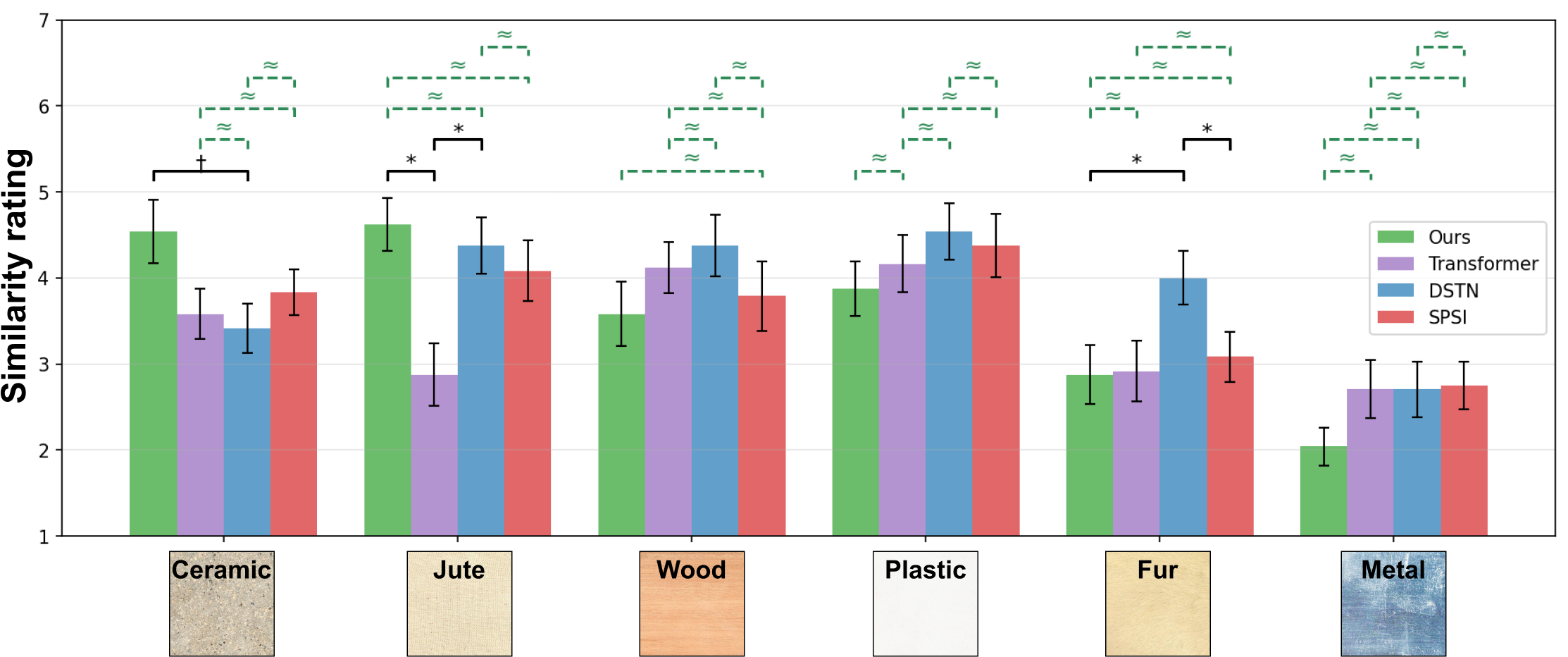}
    \caption{Perceptual similarity ratings on a 7-point Likert scale for each material and method. Each bar represents the mean and error bars indicate the standard error. Brackets indicate pairwise statistical comparisons: $\dagger$ denotes a trend toward significance ($p < 0.1$), $*$ denotes a significant difference ($p < 0.05$), and green $\approx$ denotes statistical equivalence ($p < 0.05$).}
    % JP: 各素材・各手法における7段階リッカート尺度での知覚的類似度評価。各棒は平均値を、誤差棒は標準誤差を示す。括弧は対比較の統計的比較を示し、$\dagger$ は有意傾向（$p < 0.1$）、$*$ は有意差（$p < 0.05$）、緑の $\approx$ は統計的同等性（$p < 0.05$）を表す。
    \label{fig:similarity}
\end{figure*}

\subsection{Comparative Study on Haptic Perceptual Similarity}
\label{sec:study-similarity}

This study evaluated the perceptual similarity between haptic stimuli generated by the proposed and baseline methods and those produced by real physical objects, using the VR haptic system shown in \cref{fig:VR_system}.
% JP: 本実験では、\cref{fig:VR_system} に示すVR触覚システムを用いて、提案手法およびベースライン手法により生成された触覚刺激と実物体から生じる触覚刺激との知覚的類似度を評価した。

\subsubsection{Apparatus and Task}

As shown in \cref{fig:similarity_setup}, a real object and its virtual counterpart were placed side by side in the experimental space.
Participants alternately traced the real and virtual objects using the stylus pen, and rated the perceptual similarity of the haptic sensations on a 7-point Likert scale in response to the question ``How similar are these objects?'' (1: Not similar at all, 7: Extremely similar).
Six materials were selected from those shown in \cref{fig:materials} (Ceramic, Jute, Wood, Plastic, Fur, and Metal), and the haptic stimuli for each material were rendered using all four methods evaluated in the technical evaluation.
This study specifically targets the perceptual impact of waveform quality itself, complementing the latency perception study (Sec.~\ref{sec:study-latency}). To isolate waveform quality as the sole variable under evaluation, the total end-to-end system latency was equalized across all methods by matching the inference time of the Transformer, which had the longest inference latency among the evaluated methods.
To avoid visual bias, the real objects were occluded from participants' view throughout the experiment, the identity of the material being touched was not disclosed, and the virtual objects in VR space were rendered with a uniform, material-agnostic appearance.
% JP: \cref{fig:similarity_setup} に示すように、実験空間内に実物体とその仮想対応物を並置した。
% JP: 参加者はスタイラスペンを用いて実物体と仮想物体を交互になぞり、「これらの物体はどの程度似ていますか？」（1: 全く似ていない、7: 非常に似ている）という質問に対し、触覚の知覚的類似度を7段階リッカート尺度で評価した。
% JP: \cref{fig:materials} に示す素材から6種（Ceramic、Jute、Wood、Plastic、Fur、Metal）を選択し、各素材の触覚刺激は技術評価で用いた4手法すべてで生成した。
% JP: 波形品質を唯一の評価変数として分離するため、評価対象手法の中で最も推論レイテンシが長い Transformer の推論時間に合わせることで、すべての手法間でエンドツーエンドのシステムレイテンシを統一した。
% JP: 視覚的バイアスを排除するため、実験中は実物体を参加者の視界から遮蔽し、触れている素材の正体は開示せず、VR空間内の仮想物体は素材に依存しない均一な外観で描画した。

\subsubsection{Participants and Procedure}

The same 24 participants from the latency perception study (\cref{sec:study-latency}) took part, wearing the same set of devices as before.
% JP: レイテンシ知覚実験（\cref{sec:study-latency}）と同じ24名の参加者が、同一のデバイス一式を装着して参加した。

Participants responded to a total of 24 conditions (6 materials $\times$ 4 methods).
All four methods were presented consecutively for a given material before moving on to the next, and the order of materials and the order of methods within each material were counterbalanced and randomized across participants.
A rest break was provided after every 12 conditions.
% JP: 参加者は合計24条件（6素材 $\times$ 4手法）に回答した。
% JP: 各素材について4手法を連続で提示した後に次の素材に移行し、素材の順序および各素材内の手法の順序は参加者間でカウンターバランスおよびランダム化した。
% JP: 12条件ごとに休憩を設けた。

At the start of each condition, participants first traced all four edges of the real object followed by the virtual object in that order.
They then freely explored both objects and submitted their rating at a time of their choosing.
Participants were instructed to base their similarity judgments on the vibratory sensation transmitted through the stylus pen.
% JP: 各条件の開始時、参加者はまず実物体の4辺すべてをなぞり、続いて仮想物体を同じ順序でなぞった。
% JP: その後、両物体を自由に探索し、任意のタイミングで評価を提出した。
% JP: 参加者には、スタイラスペンを通じて伝達される振動感覚に基づいて類似度を判断するよう教示した。

\subsubsection{Results}
% fig~\ref{fig:similarity}に各マテリアル、各手法ごとの触感の類似性の平均値と標準誤差を示す。
% 図中のアスタリスクは有意差（*:p<0.05）、ダガーは有意傾向（†:p<0.1）、アプロキシメイト（≈:p<0.05）は同等性を示している。
% また、それぞれの数値をtable~\ref{tab:similarity}に示す。
\Cref{fig:similarity} shows the mean and standard error of haptic perceptual similarity ratings for each material and method.
Asterisks in the figure indicate significant differences ($*$: $p < 0.05$), daggers indicate marginal significance ($\dagger$: $p < 0.1$), and the approximately-equal symbol ($\approx$) indicates statistical equivalence ($p < 0.05$).
% JP: \Cref{fig:similarity} に各素材・各手法における触覚知覚類似度評価の平均値と標準誤差を示す。
% JP: 図中のアスタリスクは有意差（$*$: $p < 0.05$）、ダガーは有意傾向（$\dagger$: $p < 0.1$）、近似等号記号（$\approx$）は統計的同等性（$p < 0.05$）を示す。

We first assessed the normality of each data distribution using the Shapiro-Wilk test, which rejected normality for all materials ($p < 0.05$).
We therefore applied the non-parametric Friedman test, which revealed significant differences among methods for Ceramic, Jute, and Fur ($p < 0.05$).
For materials showing significant differences, we conducted post-hoc pairwise comparisons using the Wilcoxon signed-rank test with Shaffer correction.
The proposed method received significantly higher ratings than Transformer for Jute ($p = 0.013$), and showed a marginal trend toward higher ratings than DSTN for Ceramic ($p = 0.075$).
Conversely, the proposed method received significantly lower ratings than DSTN for Fur ($p =  0.047$).
% JP: まず Shapiro-Wilk 検定により各データ分布の正規性を評価したところ、すべての素材で正規性が棄却された（$p < 0.05$）。
% JP: そのため、ノンパラメトリックな Friedman 検定を適用した結果、Ceramic、Jute、Fur において手法間に有意差が認められた（$p < 0.05$）。
% JP: 有意差が認められた素材について、Shaffer 補正を伴う Wilcoxon 符号順位検定による事後多重比較を実施した。
% JP: 提案手法は Jute において Transformer より有意に高い評価を受け（$p = 0.013$）、Ceramic において DSTN より高い評価への有意傾向を示した（$p = 0.075$）。
% JP: 一方、Fur においては提案手法は DSTN より有意に低い評価を受けた（$p = 0.047$）。

% また、手法間の同等性を合わせて調べるために、Two One-Sided Tests（Wilcoxon signed-rank test）を行った。〇〇のような条件下で同等性の検定を行った。
% 結果、多くの手法間において同等性が示された（p<0.05）。
To further examine equivalence between methods, we performed Two One-Sided Tests using the Wilcoxon signed-rank test with an equivalence bound of $\pm\delta = 0.8 \times \text{pooled SD}$ per pair (Cohen's $d = 0.8$, $\alpha = 0.05$). Equivalence with the proposed method was confirmed for DSTN and SPSI in Jute ($p = 0.015$, $0.019$), SPSI in Wood ($p < 0.001$), Transformer in Plastic ($p = 0.043$), Transformer and SPSI in Fur ($p < 0.001$), and Transformer and DSTN in Metal ($p = 0.038$, $0.045$).
% JP: さらに手法間の同等性を検討するため、Wilcoxon 符号順位検定を用いた Two One-Sided Tests を同等性境界 $\pm\delta = 0.8 \times \text{pooled SD}$（Cohen's $d = 0.8$, $\alpha = 0.05$）で実施した。
% JP: 提案手法との同等性は、Jute における DSTN および SPSI（$p = 0.015$, $0.019$）、Wood における SPSI（$p < 0.001$）、Plastic における Transformer（$p = 0.043$）、
% JP: Fur における Transformer および SPSI（$p < 0.001$）、Metal における Transformer および DSTN（$p = 0.038$, $0.045$）で確認された。

These results are discussed in relation to H3 in \cref{sec:discussion}.
% JP: これらの結果はH3との関連で\cref{sec:discussion}にて議論する。

\section{Discussion}
\label{sec:discussion}

% JP: （ロードマップ段落削除：サブセクション見出しで構造が自明なため）

\subsection{H1: Waveform Reproduction and Inference Latency}

H1 is supported. As shown in \cref{tab:results}, the proposed model achieves the highest GFC (0.96) and lowest RMSE (0.22) among all methods, while simultaneously attaining the shortest inference latency (5.2\,ms). These results demonstrate that the Flow Matching-based generative approach achieves superior waveform reproduction accuracy and lower inference latency compared to existing baseline methods trained with deterministic reconstruction losses.
% JP: H1は支持される．\cref{tab:results}に示すように，提案モデルは全手法中最高のGFC（0.96）と最低のRMSE（0.22）を達成しつつ，最短の推論レイテンシ（5.2\,ms）を同時に実現した．これらの結果は，Flow Matchingに基づく生成的アプローチが，決定論的再構成損失で学習された既存ベースライン手法と比較して，優れた波形再現精度とより低い推論遅延を達成することを実証している．

The waveforms and frequency spectra in \cref{fig:results} provide further insight into these differences.
Transformer produces overly smooth outputs with attenuated high-frequency components, likely because its MSE-based deterministic regression causes outputs to converge toward the distributional mean when the haptic data is multimodal.
SPSI exhibits global phase shifts, suggesting that its non-iterative phase recovery fails to maintain continuity with preceding waveform segments.
DSTN approaches the proposed model in GFC (0.94 vs.\ 0.96) but requires approximately twice the inference time (11.9\,ms vs.\ 5.2\,ms).
The proposed Flow Matching-based model, as a generative approach that learns the conditional data distribution rather than performing point-estimate regression, can reproduce diverse haptic patterns without such averaging artifacts.
% JP: \cref{fig:results}に示す波形と周波数スペクトルは，これらの差異についてさらなる知見を提供する．
% JP: Transformerは高周波成分が減衰した過度に平滑な出力を生成しており，これはMSEベースの決定論的回帰が，触覚データがマルチモーダルである場合に出力を分布の平均に収束させるためと考えられる．
% JP: SPSIは全体的な位相シフトを示しており，非反復的位相復元が先行波形セグメントとの連続性を維持できていないことが示唆される．
% JP: DSTNはGFCにおいて提案モデルに近い値を示す（0.94 vs.\ 0.96）が，推論時間は約2倍を要する（11.9\,ms vs.\ 5.2\,ms）．
% JP: 提案するFlow Matchingベースのモデルは，点推定回帰ではなく条件付きデータ分布を学習する生成的アプローチとして，このような平均化アーティファクトなしに多様な触覚パターンを再現できる．

\subsection{H2: Visual-Haptic Latency Perception}

H2 is supported. The estimated perceptual thresholds (130.3\,ms for Turn-On, 107.7\,ms for Turn-Off) both exceed the total end-to-end system latency of the proposed model (approximately 22\,ms), confirming that users can interact with the VR haptic system without perceiving visual-haptic asynchrony.
% JP: H2は支持される．推定された知覚閾値---Turn-On条件で130.3\,ms，Turn-Off条件で107.7\,ms---はいずれも提案モデルのエンドツーエンド総レイテンシ（約22\,ms）を上回っており，ユーザが視覚-触覚の非同期を知覚することなくVR触覚システムとインタラクションできることが確認された．

All baseline methods also fall below the thresholds, but the proposed model provides the largest margin, offering greater headroom for future increases in model complexity or dataset scale.
This margin also suggests that the system can tolerate the network latency of wireless or cloud-based HMD connections, or the slower on-device inference of a standalone HMD, broadening the range of deployable VR configurations.

% JP: 全ベースライン手法も閾値を下回るが，提案モデルは最大のマージン（Turn-Off閾値に対して約85\,ms，Transformerは77\,ms）を提供し，将来のモデル複雑化やデータセット拡大に対するより大きな余裕を与える．
% JP: このマージンは，無線接続やクラウドベースのHMD接続によるネットワーク遅延，あるいはスタンドアローンHMDにおけるより低速なオンデバイス推論にも耐えうることを示唆しており，展開可能なVR構成の範囲を広げるものである．

Our thresholds are lower than the 160\,ms reported for a tabletop display with a stylus pen~\cite{HaptoMapping2023}, possibly because a VR HMD replaces the entire visual field, heightening sensitivity to visual-haptic asynchrony. These thresholds can serve as a practical latency budget for future vibrotactile generative models targeting VR HMD environments.
% JP: 本閾値は卓上ディスプレイとスタイラスペンで報告された160\,ms~\cite{HaptoMapping2023}より低く，VR HMDが視野全体を置換するため視覚-触覚非同期への感度が高まる可能性がある．これらの閾値は，VR HMD環境を対象とした将来の振動触覚生成モデルのための実用的なレイテンシバジェットとなり得る．

\subsection{H3: Haptic Perceptual Similarity}

H3 is partially supported. The proposed method achieved significantly higher ratings for Jute (vs.\ Transformer, $p = 0.013$) and showed a marginal trend for Ceramic (vs.\ DSTN, $p = 0.075$). Both are hard materials with high surface roughness that produce vibrotactile signals with prominent spectral peaks. As shown in \cref{fig:results}~(b), the proposed method reproduces these dominant spectral components more faithfully than the baselines. Because roughness perception is closely linked to the spectral content of vibrotactile stimuli~\cite{VibrotactileRoughness2024, VisuoTactileRoughness2025}, this superior spectral reproduction likely contributed to the higher perceived similarity.
% JP: H3は部分的に支持される．提案手法はJute（vs.\ Transformer，$p = 0.013$）において有意に高い評価を獲得し，Ceramic（vs.\ DSTN，$p = 0.075$）においても有意傾向を示した．これらはいずれも表面粗さが大きい硬質素材であり，顕著なスペクトルピークを持つ振動触覚信号を生成する．\cref{fig:results}~(b)に示すように，提案手法はこれらの支配的なスペクトル成分をベースライン手法よりも忠実に再現している．粗さ知覚は振動触覚刺激のスペクトル内容と密接に関連しているため~\cite{VibrotactileRoughness2024, VisuoTactileRoughness2025}，この優れたスペクトル再現性が知覚的類似度の向上に寄与したと考えられる．

Conversely, for Fur, the proposed method was rated lower than DSTN ($p = 0.047$); however, all methods received uniformly low ratings (approximately 2--3), likely because Fur's tactile sensation is dominated by compliance cues that vibrotactile feedback alone cannot convey.
% JP: 一方，Furについては提案手法はDSTNより低い評価を受けた（$p = 0.047$）が，すべての手法で評価値は均一に低く（約2--3），Furの触感はコンプライアンス手がかりが支配的であり振動触覚フィードバックのみでは伝達困難であるためと考えられる．

For all materials except Ceramic, equivalence with one or more baselines was confirmed via TOST, suggesting that the waveform accuracy differences among methods do not exceed human perceptual thresholds for these materials. The limited diversity of training data and the finite bandwidth of the haptic actuator may also have constrained the perceptual fidelity attainable by any method.
% JP: Ceramicを除くすべての素材において，TOSTにより1つ以上のベースライン手法との同等性が確認され，手法間の波形精度の差異がこれらの素材に対する人間の知覚閾値を超えていないことが示唆された．訓練データの多様性の限界やハプティックアクチュエータの帯域幅の制約も，いずれの手法で達成可能な知覚的忠実度を制限していた可能性がある．

\section{Limitations and Future Work}
\subsection{Limitations in Haptic Expressiveness}
Although the model achieved high waveform reproduction accuracy in the technical evaluation, the perceptual similarity ratings in the user study were more modest. We attribute this gap to factors outside the generative model: the training data and the physical constraints of the haptic actuator.
% JP: 提案モデルは技術的評価において高い波形再現精度を達成したが，ユーザスタディでの知覚的類似度評価はより控えめであった．このギャップの原因は，生成モデル自体の外にある2つの要因，すなわち学習データと触覚アクチュエータの物理的制約に帰する．

The first factor is the training data. The model faithfully reproduces the training waveforms (GFC of 0.96), but the Cluster Haptic Texture Dataset~\cite{ClusterHapticTextureDataset2025} was recorded by sliding an artificial finger on a numerically controlled machine at fixed speeds and forces, so its distribution may not cover the diversity of unconstrained human interaction. Collecting more naturalistic data over a wider range of interaction conditions (e.g., the stylus contact angle) and materials is a broadly applicable route to better perceptual quality.
% JP: 第1の要因は学習データである．モデルは学習波形を忠実に再現するが（GFC 0.96），学習データであるCluster Haptic Texture Dataset~\cite{ClusterHapticTextureDataset2025}は数値制御マシンに取り付けた人工指を一定の速度・力で素材上をなぞらせて収録したものであり，その分布は非制約的な人間のインタラクションの多様性をカバーしていない可能性がある．より広範なインタラクション条件（例：スタイラスの接触角）や素材にわたる，より自然な（naturalistic）データの収集は，知覚品質を高めるための広く適用可能な方向性である．

The second factor is the haptic actuator (Haptic Reactor), whose finite frequency response makes the perceived waveform a filtered rendition of the model output. Its resonance peaks near 160 and 320 Hz, adopted in prior VR haptic research (e.g., HaptoFloater~\cite{HaptoFloater2024}), cover most of the generated content, but the response is not flat: the 50--100 Hz region is attenuated, and low-amplitude texture cues can fall below its dynamic range. Broader-bandwidth hardware, or an actuator model folded into the pipeline to compensate at inference, would narrow this gap.
% JP: 第2の要因は触覚アクチュエータ（Haptic Reactor）であり，その有限な周波数応答のため，知覚される波形はモデル出力のフィルタリングされた再現となる．160 Hzおよび320 Hz付近の共振ピーク（先行するVR触覚研究，例：HaptoFloater~\cite{HaptoFloater2024}でも採用）は生成波形の多くをカバーするが，応答は平坦ではなく，50--100 Hz帯域は減衰し，低振幅のテクスチャ手がかりはダイナミックレンジを下回りうる．このギャップは，より広帯域のハードウェアか，推論時に補償するアクチュエータモデルのパイプラインへの組み込みによって縮小できる．

The setup is also single-axis: following prior work, the pipeline collapses the three-axis acceleration via DFT321~\cite{DFT321}, whose perceptual cost is unexamined. Multi-axis actuation would render the lateral shear force, influential for soft, high-friction materials such as Fur, and let this reduction loss be assessed, both of which we leave to future work.
% JP: 構成は1軸でもある：本研究のパイプラインは既存研究に倣い3軸加速度をDFT321~\cite{DFT321}で圧縮しており，その知覚的損失は未検証である．マルチ軸アクチュエーションは，Furのような柔らかく摩擦の高い素材で影響の大きい横方向のせん断力を提示し，この圧縮による損失の評価も可能にする．いずれも今後の課題とする．

\subsection{Generalization and Scalability of the Model}
In its current form, the proposed model accepts a discrete material label, restricting haptic generation to materials seen during training. This reflects the primary focus of this work on signal fidelity and inference speed rather than broadening the conditioning interface. Future work can attach front-end modules that map richer inputs (such as text or images~\cite{GANHaptics2021,HapticGen2025}) onto the material embedding space while keeping the trained waveform generation backbone intact.
% JP: 現在の形態では，提案モデルは条件付け入力の一部として離散的な素材ラベルを受け付けるため，触覚生成は学習時に見た素材に限定される．この制限は，条件付けインターフェースの拡張よりも，信号忠実度と推論速度の観点から触覚波形生成の基盤的能力を向上させることに主眼を置いた本研究の焦点に起因するものである．この制限を克服するために，今後の研究では，より豊富な入力モダリティ（自然言語による記述や表面テクスチャの画像など）を提案モデルが既に使用している素材埋め込み空間にマッピングする専用のフロントエンドモジュールを追加することでモデルを拡張できる．先行研究ではテキストおよび視覚入力を条件とした触覚生成の実現可能性が実証されており~\cite{GANHaptics2021,HapticGen2025}，それらの条件付けモジュールは，学習済みの波形生成バックボーンを維持したまま提案モデルに統合可能である．

Regarding the output modality, the current model generates vibrotactile waveforms exclusively, reflecting the broader state of the field where the majority of existing haptic generative models and datasets are likewise confined to vibrotactile stimuli. Extending the output to other haptic channels (such as force feedback or thermal stimulation) would first require collecting dedicated training data, as no sufficient datasets currently exist. However, because the model operates on time-series representations in a latent space, it can in principle be applied to any haptic modality expressible as a continuous time-series waveform, provided that an appropriate signal encoder--decoder is trained for that modality.
% JP: 出力モダリティに関しては，現在のモデルは振動触覚波形のみを生成するが，これは当該分野の現状を反映したものである：既存の触覚生成モデルおよび公開データセットの大多数も同様に振動触覚刺激に限定されている．力覚フィードバックや温度刺激などの他の触覚チャネルへの出力拡張には，まず各モダリティの専用学習データの収集が必要となるが，十分なデータセットは現時点では存在しない．しかしながら，モデリングの観点からは，提案モデルのアーキテクチャは本質的に振動信号に限定されるものではない．モデルが潜在空間における時系列表現上で動作するため，原理的には，当該モダリティに対する適切な信号エンコーダ・デコーダが学習されていれば，連続時系列波形として表現可能な任意の触覚モダリティに適用可能である．

Finally, the present study validated the model on 10 material categories, establishing the foundation for scalable haptic design. However, verifying actual scalability to a substantially larger and more diverse material set requires further expansion of the training data and remains an important direction for future work.
% JP: 本研究では10種類の素材カテゴリでモデルを検証し，スケーラブルな触覚デザインの基盤を示した．大幅に多様な素材セットへの実際のスケーラビリティの検証には訓練データのさらなる拡充が必要であり，今後の重要な研究課題である．

\section{Applications}
\begin{figure}[t]
  \centering
  \includegraphics[width=1\linewidth]{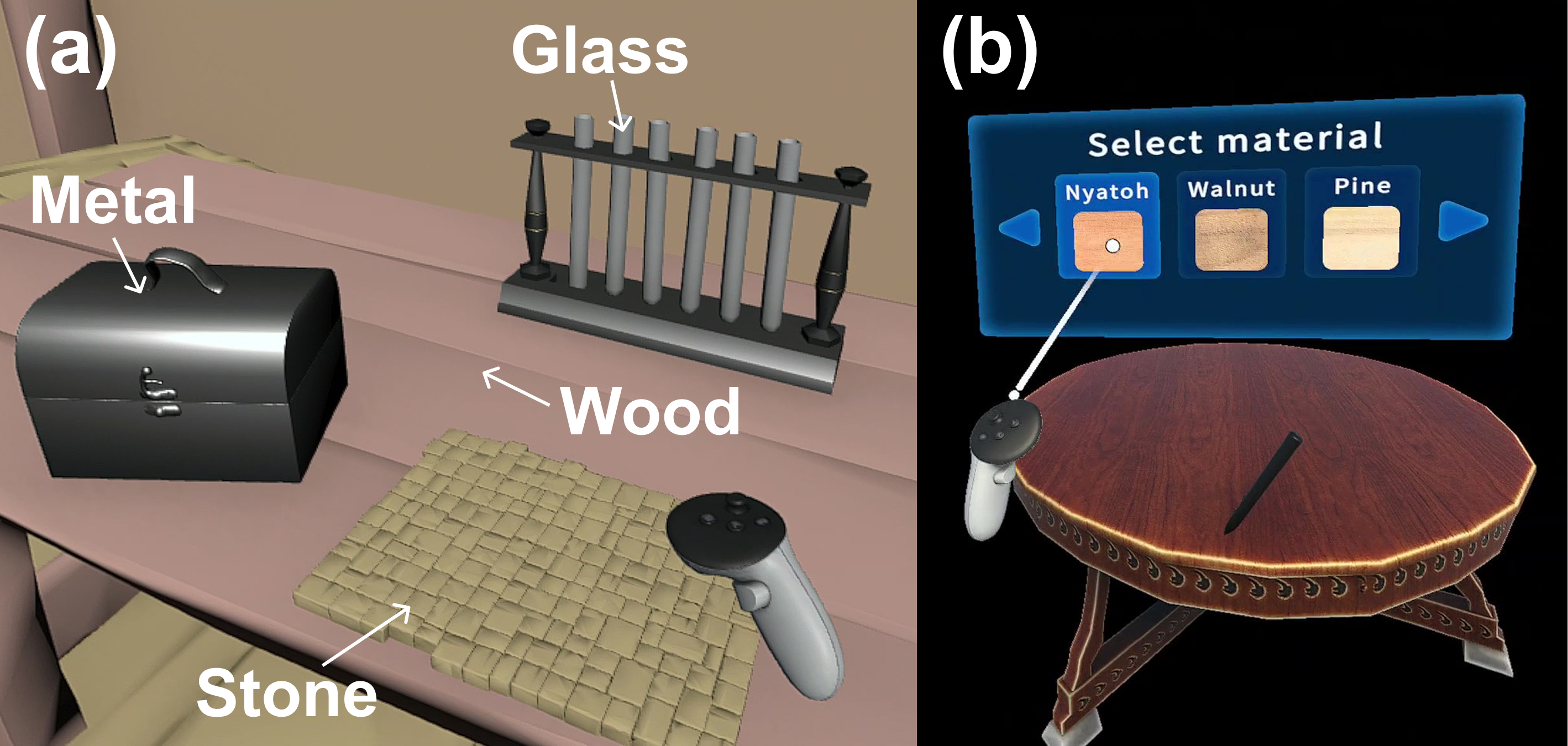}
  \caption{Application scenarios of the VR system integrated with the proposed model. (a) Scalable haptic authoring: material labels assigned to virtual objects drive on-the-fly haptic generation. (b) Surface texture design support: designers evaluate tactile feel of different materials in VR without physical prototyping.}
% JP: 提案モデルを統合したVRシステムの応用シナリオ．(a) スケーラブルな触覚オーサリング：仮想オブジェクトに割り当てられた素材ラベルがオンザフライの触覚生成を駆動する．
% JP: (b) 表面テクスチャデザイン支援：デザイナーが物理的なプロトタイピングなしにVR内で異なる素材の触感を評価する．
  \label{fig:applications}
\end{figure}

\subsection{Scalable Haptic Authoring for VR Environments}
The proposed model substantially lowers the barrier to introducing haptic feedback into VR environments (\cref{fig:applications} (a)).
By assigning a material label to each virtual object and passing interaction parameters to the model, developers can generate contextually appropriate haptic waveforms on the fly without manual waveform design for each material.
% JP: 提案モデルは，VR環境への触覚フィードバック導入の障壁を大幅に低減する（\cref{fig:applications} (a)）．
% JP: 各仮想オブジェクトに素材ラベルを割り当て，インタラクションパラメータをモデルに渡すことで，
% JP: 開発者は素材ごとに手動で波形を設計することなく，文脈に適した触覚波形をオンザフライで生成できる．

The model is also not limited to the stylus pen hardware used in this study.
For example, the Meta Quest Touch Plus Controller supports independent control of vibration amplitude and frequency via the Meta XR Haptics SDK.
Although this interface does not permit direct output of raw waveforms, compatible control can be achieved by applying amplitude-modulation-based actuation methods~\cite{ISM_WHC2021,ISM_MetaQuest2026}, enabling the proposed model to drive such devices.
% JP: 本モデルは，本研究で使用したスタイラスペンのハードウェアに限定されるものではない．
% JP: 例えば，Meta Quest Touch Plus ControllerはMeta XR Haptics SDKを通じて振動の振幅と周波数の独立制御に対応している．
% JP: このインターフェースは生の波形の直接出力を許可しないが，振幅変調ベースの駆動手法~\cite{ISM_WHC2021,ISM_MetaQuest2026}を適用することで
% JP: 互換的な制御が実現可能であり，提案モデルによるこれらのデバイスの駆動が可能となる．

\subsection{Support for Surface Texture Design}
The ability to generate haptic waveforms conditioned on diverse material labels makes the proposed model well-suited for assisting in the surface texture design of physical products.
As shown in \cref{fig:applications}~(b), a designer can iteratively explore different surface textures in a virtual environment by querying the model with target material labels, and experience the resulting haptic feedback in real time.
This workflow enables designers to evaluate the tactile feel of a product surface during the design process itself, reducing the need for costly physical prototyping and allowing more rapid iteration.
% JP: 多様な素材ラベルを条件として触覚波形を生成する能力により，提案モデルは物理デバイスの表面テクスチャデザインの支援に適している．
% JP: \cref{fig:applications}~(b)に示すように，デザイナーは仮想環境内で目標とする素材ラベルをモデルに入力し，
% JP: 異なる表面テクスチャを反復的に探索しながら，その結果得られる触覚フィードバックをリアルタイムで体験できる．
% JP: このワークフローにより，デザイナーはデザインプロセス自体の中でデバイス表面の触感を評価でき，
% JP: 高コストな物理プロトタイピングの必要性を低減し，より迅速な反復を可能にする．

\section{Conclusion}

This paper presented HaptoFlow, a vibrotactile generative model based on Flow Matching designed for real-time haptic rendering in Virtual Reality.
By learning a continuous vector field that efficiently transports a noise distribution toward the target waveform distribution, HaptoFlow simultaneously achieves high waveform reproduction accuracy and low inference latency conditioned on material labels and interaction parameters.
% JP: 本論文では，Virtual Realityにおけるリアルタイム触覚レンダリングのために設計された，Flow Matchingに基づく振動触覚生成モデルHaptoFlowを提示した．
% JP: ノイズ分布を目標波形分布へ効率的に輸送する連続ベクトル場を学習することで，HaptoFlowは素材ラベルおよびインタラクションパラメータを条件として，高い波形再現精度と低い推論遅延の両立を達成する．

Three experiments validated the system from complementary perspectives.
The technical evaluation confirmed that HaptoFlow outperforms all baseline methods in both waveform reproduction accuracy and inference latency.
The latency perception study established that the total end-to-end system latency falls well below the visual-haptic delay thresholds measured for a VR HMD with a stylus pen, 130.3\,ms for Turn-On and 107.7\,ms for Turn-Off.
The perceptual similarity study revealed that the proposed method yields higher similarity ratings for hard, high-roughness materials whose tactile sensations are dominated by spectral vibration cues.
% JP: 3つの実験により，システムを相補的な観点から検証した．
% JP: 技術評価では，HaptoFlowが波形再現精度と推論遅延の双方において全ベースライン手法を上回ることが確認された（H1）．
% JP: 遅延知覚実験では，システム全体のエンドツーエンドレイテンシが，VR HMDとスタイラスペンにおけるvisual-haptic遅延閾値---Turn-Onで130.3\,ms，Turn-Offで107.7\,ms---を十分に下回ることが確立された（H2）．
% JP: 知覚的類似度実験では，スペクトル振動手がかりが触覚を支配する硬質・高粗さ素材において，提案手法がより高い類似度評価を得ることが明らかになった（H3，部分的に支持）．

The gap between high signal-level accuracy and moderate perceptual ratings points to limitations in training data diversity and actuator bandwidth rather than in the generative framework itself, suggesting clear avenues for improvement.
Future work includes expanding the training data to cover broader materials and interaction conditions, incorporating actuator-aware generation to compensate for hardware limitations, and extending the conditioning interface to accept richer input modalities such as natural language and images.
% JP: 高い信号レベルの精度と控えめな知覚評価との乖離は，生成フレームワーク自体ではなく，学習データの多様性とアクチュエータの帯域幅における限界を指し示しており，明確な改善の方向性を示唆している．
% JP: 今後の課題として，より広範な素材やインタラクション条件を網羅する学習データの拡充，ハードウェア制約を補償するアクチュエータ対応型生成の導入，および自然言語や画像といったより豊富な入力モダリティを受け付ける条件付けインターフェースの拡張が挙げられる．

\acknowledgments{This study was supported by JST ACT-X Grant Number JPMJAX25C4 and JSPS KAKENHI Grant Number JP25H00722, Japan.}

\bibliographystyle{abbrv-doi}

\bibliography{ismar2026}
\end{document}